\documentclass{aastex631}

\usepackage{amsmath}
\usepackage{hyperref}

\usepackage{graphicx}
\usepackage{subfigure}

\usepackage{CJK}

\usepackage{xcolor}

\begin{document}

\title{Random Polarization Position Angle Behaviors across Bursts of Repeating Fast Radio Bursts}

\begin{CJK*}{UTF8}{gbsn}
\author[0000-0002-2552-7277]{Xiaohui Liu (刘小辉)}
\affiliation{State Key Laboratory of Radio Astronomy and Technology, National Astronomical Observatories, CAS, A20 Datun Road, Chaoyang District, Beijing, 100101, P. R. China}
\affiliation{School of Astronomy and Space Science, University of Chinese Academy of Sciences, Beijing 100049, People's Republic of China}

\author[0000-0001-8065-4191]{Jiarui Niu (牛佳瑞)}
\affiliation{National Astronomical Observatories, Chinese Academy of Sciences, 20A Datun Road, Chaoyang District, Beijing 100101, People's Republic of China}

\author[0000-0002-9332-5562]{Tiancong Wang (王天聪)}
\affiliation{Department of Astronomy, Beijing Normal University, Beijing 100875, People's Republic of China}

\author[0009-0005-8586-3001]{Jun-Shuo Zhang (张钧硕)}
\affiliation{National Astronomical Observatories, Chinese Academy of Sciences, 20A Datun Road, Chaoyang District, Beijing 100101, People's Republic of China}
\affiliation{School of Astronomy and Space Science, University of Chinese Academy of Sciences, Beijing 100049, People's Republic of China} 

\author[0000-0003-4721-4869]{Yuanhong Qu (屈元鸿)}
\affiliation{The Nevada Center for Astrophysics, University of Nevada, Las Vegas, NV 89154, USA}
\affiliation{Department of Physics and Astronomy, University of Nevada, Las Vegas, NV 89154, USA}

\author[0000-0002-6465-0091]{Jinchen Jiang (姜金辰)}
\affiliation{National Astronomical Observatories, Chinese Academy of Sciences, 20A Datun Road, Chaoyang District, Beijing 100101, People's Republic of China}

\author[0000-0002-8744-3546]{Yongkun Zhang (张永坤)}
\affiliation{National Astronomical Observatories, Chinese Academy of Sciences, 20A Datun Road, Chaoyang District, Beijing 100101, People's Republic of China}
\affiliation{School of Astronomy and Space Science, University of Chinese Academy of Sciences, Beijing 100049, People's Republic of China} 

\author[0000-0002-5031-8098]{Heng Xu
(胥恒)}
\affiliation{National Astronomical Observatories, Chinese Academy of Sciences, 20A Datun Road, Chaoyang District, Beijing 100101, People's Republic of China}

\author[0000-0002-6423-6106]{Dejiang Zhou (周德江)}
\affiliation{National Astronomical Observatories, Chinese Academy of Sciences, 20A Datun Road, Chaoyang District, Beijing 100101, People's Republic of China}

\author[0000-0001-9036-8543]{Wei-Yang Wang (王维扬)}
\affiliation{School of Astronomy and Space Science, University of Chinese Academy of Sciences, Beijing 100049, People's Republic of China}
\email{wywang@ucas.ac.cn}

\author[0000-0001-5105-4058]{Weiwei Zhu (朱炜玮)} 
\affiliation{National Astronomical Observatories, Chinese Academy of Sciences, 20A Datun Road, Chaoyang District, Beijing 100101, People's Republic of China}

\author[0000-0002-9725-2524]{Bing Zhang (张冰)}
\affiliation{The Nevada Center for Astrophysics, University of Nevada, Las Vegas, NV 89154, USA}
\affiliation{Department of Physics and Astronomy, University of Nevada, Las Vegas, NV 89154, USA}
\email{bing.zhang@unlv.edu}

\author[0000-0001-6475-8863]{Xuelei Chen (陈学雷)}
\affiliation{State Key Laboratory of Radio Astronomy and Technology, National Astronomical Observatories, CAS, A20 Datun Road, Chaoyang District, Beijing, 100101, P. R. China}
\affiliation{School of Astronomy and Space Science, University of Chinese Academy of Sciences, Beijing 100049, People's Republic of China}
\email{xuelei@cosmology.bao.ac.cn}

\author[0000-0002-6165-0977]{Xiang-Han Cui (崔翔翰)}
\affiliation{National Astronomical Observatories, Chinese Academy of Sciences, Beijing 100101, China}
\affiliation{ASTRON, the Netherlands Institute for Radio Astronomy, Oude Hoogeveensedijk 4,7991 PD Dwingeloo, The Netherlands}

\author[0000-0002-9274-3092]{Jinlin Han (韩金林)} 
\affiliation{National Astronomical Observatories, Chinese Academy of Sciences, 20A Datun Road, Chaoyang District, Beijing 100101, People's Republic of China}

\author[0000-0002-1435-0883]{Kejia Lee (李柯伽)} 
\affiliation{National Astronomical Observatories, Chinese Academy of Sciences, 20A Datun Road, Chaoyang District, Beijing 100101, People's Republic of China}
\affiliation{Kavli Institute for Astronomy and Astrophysics, Peking University, Beijing 100871, People's Republic of China}
\affiliation{Department of Astronomy, Peking University, Beijing 100871, People's Republic of China}

\author[0000-0003-3010-7661]{Di Li
\begin{CJK}{UTF8}{bsmi}
(李菂)
\end{CJK}}
\affiliation{New Cornerstone Science Laboratory, Department of Astronomy, Tsinghua University, Beijing 100084, People's Republic of China}
\affiliation{National Astronomical Observatories, Chinese Academy of Sciences, 20A Datun Road, Chaoyang District, Beijing 100101, People's Republic of China}

\author[0000-0002-9642-9682]{Jia-Wei Luo (罗佳伟)}
\affiliation{College of Physics and Hebei Key Laboratory of Photophysics Research and Application, Hebei Normal University, Shijiazhuang, Hebei 050024, People's Republic of China}
\affiliation{Shijiazhuang Key Laboratory of Astronomy and Space Science, Hebei Normal University, Shijiazhuang, Hebei 050024, People's Republic of China}

\author[0000-0002-4300-121X]{Rui Luo (罗睿)}
\affiliation{Department of Astronomy, School of Physics and Materials Science, Guangzhou University, Guangzhou 510006, People's Republic of China}

\author[0009-0008-7428-1665]{Chengwei Liang (梁城玮)}
\affiliation{Department of Astronomy, School of Physics and Materials Science, Guangzhou University, Guangzhou, 510006, China}

\author[0000-0001-6651-7799]{Chenhui Niu (牛晨辉)}
\affiliation{Institute of Astrophysics, Central China Normal University, Wuhan 430079, People's Republic of China }

\author[0000-0001-5950-7170]{Wan-Peng Sun (孙万鹏)}
\affiliation{Liaoning Key Laboratory of Cosmology and Astrophysics, College of Sciences, Northeastern University, Shenyang 110819, China}

\author[0000-0002-9434-4773]{Bojun Wang (王铂钧)}
\affiliation{National Astronomical Observatories, Chinese Academy of Sciences, 20A Datun Road, Chaoyang District, Beijing 100101, People's Republic of China}

\author[0000-0003-4157-7714]{Fayin Wang (王发印)}
\affiliation{School of Astronomy and Space Science, Nanjing University, Nanjing 210093, People's Republic of China}
\affiliation{Key Laboratory of Modern Astronomy and Astrophysics (Nanjing University), Ministry of Education, People's Republic of China}

\author[0000-0002-3386-7159]{Pei Wang (王培)}
\affiliation{National Astronomical Observatories, Chinese Academy of Sciences, 20A Datun Road, Chaoyang District, Beijing 100101, People's Republic of China}

\author[0000-0001-6021-5933]{Qin Wu (吴沁)}
\affiliation{School of Astronomy and Space Science, Nanjing University, Nanjing 210093, People's Republic of China}
\affiliation{Key Laboratory of Modern Astronomy and Astrophysics (Nanjing University), Ministry of Education, People's Republic of China}

\author[0000-0002-1381-7859]{Ziwei Wu (吴子为)}
\affiliation{National Astronomical Observatories, Chinese Academy of Sciences, 20A Datun Road, Chaoyang District, Beijing 100101, People's Republic of China}

\author[0009-0008-7805-091X]{Jiangwei Xu (徐江伟)}
\affiliation{National Astronomical Observatories, Chinese Academy of Sciences, 20A Datun Road, Chaoyang District, Beijing 100101, People's Republic of China}

\author[0000-0001-6374-8313]{Yuan-Pei Yang (杨元培)} 
\affiliation{South-Western Institute for Astronomy Research, Yunnan University, Kunming, Yunnan 650504, People's Republic of China}
\affiliation{Purple Mountain Observatory, Chinese Academy of Sciences, Nanjing 210023, People's Republic of China} 

\author[0009-0006-1500-441X]{Shiqian Zhao (赵仕乾)}
\affiliation{Department of Astronomy, School of Physics and Materials Science, Guangzhou University, Guangzhou, 510006, China}




\begin{abstract}
Fast radio bursts (FRBs) are highly polarized and mostly show a nearly constant polarization position angle (PA) during each burst.
Their PAs are observed to vary from burst to burst, with the statistical properties remaining stable across different observation sessions.
We found that the intrinsic PAs of repeating FRBs are approximately Gaussian distributed with a stable mean PA. This suggests that the magnetic field line sampled by the emission does not vary randomly across the magnetosphere, but instead remains confined to a limited range of viewing geometries, consistent with emission arising from a localized region.
A periodicity search of the PA time series using the Lomb-Scargle periodogram reveals no credible periodic signal in the period range from 10 ms to $10^7$ ms, and similar analyses of several active observations also yield null detections. We interpret these properties by extending the rotating vector model to include a dynamically evolving magnetosphere, in which the effective magnetic axis varies from burst to burst due to stochastic perturbations. In this framework, the observed PA distributions can naturally arise from geometric projection effects, and the absence of periodicity reflects the random wandering of the magnetic axis within a confined region. This scenario provides a natural explanation for both repeating and apparently non-repeating FRBs.
\end{abstract}

\keywords{Neutron stars(1108) --- Radio bursts (1339) --- Radio transient sources (2008)}


\section{Introduction} \label{sec:intro}
Fast Radio Bursts are bright radio transients likely from the most compact stars in the universe \citep{2007Sci...318..777L, 2013Sci...341...53T, 2014ApJ...790..101S, 2022Natur.602..585K}. Their true natures remain to be uncovered \citep{2019A&ARv..27....4P, 2019ARA&A..57..417C, 2023RvMP...95c5005Z}. Magnetars are rotating neutron stars with an extraordinarily strong magnetic field \citep{2017ARA&A..55..261K} and are among the leading candidates for FRB sources \citep{2016Natur.531..202S,2019ApJ...879....4W}.
This hypothesis is strongly supported by the FRB-like bursts detected from a magnetar, SGR J1935+2154 in the Milky Way \citep{2020Natur.587...59B, 2020Natur.587...54C, 2021NatAs...5..378L, 2020Natur.587...63L, 2021NatAs...5..401T}. Notably, the bright radio bursts appear to occur at random rotational phases relative to the lower-luminosity pulses \citep{2023SciA....9F6198Z}. The magnetar origin of FRBs is further corroborated by extensive polarization observations \citep{2020Natur.586..693L, 2024ApJ...972L..20N, 2024NSRev..12E.293J, 2025Natur.637...43M, 2025ApJ...988..175L}.

A growing body of evidence from polarization studies strongly suggests a magnetospheric origin for FRBs.
Most FRBs are highly linearly polarized, which matches pulsars' radiation \citep{2022RAA....22l4003J}. Even though most repeating bursts show a nearly constant position angle (PA) behavior during each burst \citep{2018Natur.553..182M}, some bursts show significant PA swings \citep{2020Natur.586..693L}.
The pulsar-like polarization PA swing observed in an apparently non-repeating FRB, FRB 20221022A \citep{2025Natur.637...43M}, is consistent with the rotating vector model (RVM) \citep{1969ApL.....3..225R}. Similar PA swings have also been reported in active repeaters, although the inferred geometrical parameters are mutually inconsistent with each other, possibly indicating a dynamically evolving magnetic configuration \citep{2025ApJ...988..175L}.
A sudden PA orthogonal jump in FRB 20201124A can occur when the degree of linear polarization reaches a minimum \citep{2024ApJ...972L..20N}.
Similar phenomena have been widely observed in radio pulsars, and the mechanism of the orthogonal mode switch has been investigated \citep{2024ApJ...973...35M}.

The spin period of repeating FRBs is considered a key prediction of the single rotating neutron star model, as it should produce regular, pulsar-like burst periodicities. Despite accumulating thousands to tens of thousands of bursts from several active repeaters, these extensive datasets reveal no clear global spin period in the millisecond-to-second range, and their waiting time distributions usually show two peaks \citep{2019ApJ...877L..19G, 2020MNRAS.496.4565C, 2021Natur.598..267L, 2022Natur.609..685X, 2022RAA....22l4001Z, 2022RAA....22l4004N, 2022MNRAS.515.3577H, 2022MNRAS.513.1925K, 2023MNRAS.519..666J, 2023MNRAS.520.2281N, 2023ApJ...955..142Z, 2024ApJ...977..129D, 2024SciBu..69.1020Z, 2025ApJ...988...41Z, 2025arXiv250714707Z, 2026SCPMA..6949512Z, 2026MNRAS.545S2222S}. The first peak, typically occurring on millisecond scales, is usually caused by adjacent sub-bursts, while the second peak generally appears at longer timescales and serves as an indicator of the activity level of this source \citep{2022RAA....22l4001Z, 2026ApJ...998..190Q}.
The absence of a periodicity challenges the hypothesis that FRBs share the same production region, e.g., open field lines, as regular pulsations of pulsars.

In contrast, very long periods have been confirmed in several repeaters. FRB 20180916B has a $\sim$16-day period with a frequency-dependent active window \citep{2020Natur.582..351C, 2021Natur.596..505P, 2021ApJ...911L...3P, 2023ApJ...956...23S}, and FRB 20121102A has a $\sim$160-day period with a large duty cycle $> 50\%$ \citep{2020MNRAS.495.3551R, 2021MNRAS.500..448C}, as well as a claimed periodicity of $\sim$126 days for FRB 20240209A \citep{2025ApJ...983L..15P}. Furthermore, some repeaters also exhibit periodic changes in their dispersion measure (DM) or rotation measure (RM) \citep{2025arXiv250506006X, 2025ApJ...994L..32L}. These extremely long periods can be interpreted as the orbital period of binary systems \citep{2020ApJ...893L..26I, 2020MNRAS.498L...1Z, 2021ApJ...917...13S, 2021ApJ...920...54W, 2022NatCo..13.4382W, 2023A&A...673A.136R, 2025arXiv251212995S}, precession \citep{2020ApJ...892L..15Z, 2020ApJ...895L..30L, 2020MNRAS.497.1001S, 2020RAA....20..142T, 2020ApJ...893L..31Y}, regular asteroid interactions \citep{2020ApJ...895L...1D, 2021MNRAS.500.4678D, 2021MNRAS.508.2079V}, or slow rotation \citep{2020MNRAS.496.3390B}.

The absence of a detectable short-timescale period can be explained geometrically by a near-aligned configuration of the rotation and magnetic axes, which produces a large duty cycle and makes traditional time-of-arrival (TOA) based searches fail \citep{2025ApJ...988...62L, 2025ApJ...982...45B}. In such cases, PA may still encode the rotation or orbital information, as each burst could sample a small segment of an underlying periodic PA curve related to the magnetospheric geometry \citep{2025MNRAS.542L..43R}. Attempts to use PA behavior to constrain the origin of the periodicity in repeating sources, such as FRB 20180916B, have so far yielded inconclusive results \citep{2025A&A...702A.248B}, highlighting both the potential and the challenges of PA-based period searches.
The most sensitive single-dish radio telescope, the Five-hundred-meter Aperture Spherical radio Telescope (FAST) \citep{2011IJMPD..20..989N, 2018IMMag..19..112L, 2019SCPMA..6259502J}, has detected thousands upon thousands of bursts for several active repeaters and established high-quality polarization datasets, which have great potential for this PA-viewed period search problem.
In this paper, we further explore the topic of how to extract periodic information from a large sample of PA based on four large polarimetric datasets of FRB 20201124A, FRB 20220912A, and FRB 20240114A, observed by FAST under the FAST FRB Key Science Project. The four datasets consist of the first and second active episodes of FRB 20201124A, together with the datasets of FRB 20220912A and FRB 20240114A.

The rest of this paper is organized as follows.
In Section \ref{sec: SPPA}, we use the Lomb-Scargle Periodogram (LSP) to search for periodicity in the PA-time series.
In Section \ref{sec: PAb}, we show the PA population properties of our datasets.
In Section \ref{sec: sim}, we extend the traditional RVM by introducing stochastic perturbations to the magnetic axis and then apply it to our datasets.
In Section \ref{sec: imp}, we discuss some implications of our results.
Finally, we conclude the main results of this paper in Section \ref{sec: con}.

\section{Searching for periodicity in Position Angles} \label{sec: SPPA}
\subsection{Polarimetric Data of FRBs}
We take four PA samples from three active repeaters observed by the central beam of the L-band 19 beam receiver of FAST \citep{2020RAA....20...64J}, including the first and second active episodes of FRB 20201124A, the sample of FRB 20220912A, and the sample of FRB 20240114A. Following the convention of our previous work \citep{2025ApJ...988..175L}, the first and second active episodes of FRB 20201124A are referred to as FRB 20201124A1 and FRB 20201124A2, respectively. To avoid the noise-dominated false feature, the bursts with intensity signal-to-noise ratio (SNR) less than 50 are discarded in our final time-PA series. Most of the bursts have high linear polarization degrees, so the Faraday rotation can be precisely determined and removed.

The continuous PA curves occasionally exhibit orthogonal jumps \citep{2024ApJ...972L..20N}, sudden and discontinuous changes on millisecond timescales. This phenomenon is likely attributable either to the magnetospheric geometry \citep{1976Natur.263..202B, 1978ApJ...223..961C, 2026ApJ...997...37Q}, the effects of plasma lensing \citep{2026arXiv260111122L}, or the shotnoise of microstructure \citep{2026arXiv260119254B}.
Although such events break the continuity of the PA evolution and complicate the detection of periodicity, their extremely low occurrence rate suggests that they are unlikely to dominate the statistical behavior of the PA sequence. We therefore retain these jump features in the data to preserve the intrinsic PA behavior.

The data used in this work primarily consist of the time and PA of bursts, which are derived from the raw data after calibration, standard dedispersion, and correction for Faraday rotation. The detailed processing pipelines can be found in the corresponding literature \citep{2022Natur.609..685X, 2022RAA....22l4003J, 2023ApJ...955..142Z, Wang:2026ssk}.
For most astrophysical contexts, the interstellar medium is sufficiently tenuous that free-free absorption is negligible, and its magnetic field is sufficiently weak that Faraday conversion can be disregarded \citep{2025ApJ...988..164W}.
In that case, Faraday rotation dominates the propagation, in which the corresponding observed PA is given by
\begin{equation}
    \mathrm{PA_{obs}} (\lambda^2) = \mathrm{PA}_0 + \mathrm{RM}\cdot \lambda^2,
\end{equation}
where $\mathrm{PA}_0$ is the intrinsic PA.
$\rm PA_0$ characterizes the orientation of the linear polarization vector projected onto the plane of the sky, which can be used to probe the geometry of the magnetic configuration. For the scenario of a dipolar magnetic configuration, the intrinsic PA can be described by the RVM, which has been widely used in modeling the PA of radio pulsars \citep{2023MNRAS.520.4801J, 2023MNRAS.520.4582P, 2023RAA....23j4002W}.

From the de-rotated linear Stokes parameters $Q_{\mathrm{obs}, \mathrm{derot}}$ and $U_{\mathrm{obs}, \mathrm{derot}}$, the intrinsic PA can be determined using the formula,
\begin{equation}
    \mathrm{PA}_{\mathrm{obs}} = \frac{1}{2} \arctan \left( \frac{U_{\mathrm{obs}, \mathrm{derot}} }{Q_{\mathrm{obs}, \mathrm{derot}}} \right),
\end{equation}
and the uncertainty in PA can be calculated using standard error propagation:
\begin{equation}
\sigma_{\mathrm{PA}} = \frac{1}{2} \frac{Q_{\mathrm{obs}, \mathrm{derot}} U_{\mathrm{obs}, \mathrm{derot}}}{Q_{\mathrm{obs}, \mathrm{derot}}^2+U_{\mathrm{obs}, \mathrm{derot}}^2} \sqrt{ \left( \frac{\sigma_{Q_{\mathrm{obs}, \mathrm{derot}}}}{Q_{\mathrm{obs}, \mathrm{derot}}} \right)^2 + \left( \frac{\sigma_{U_{\mathrm{obs}, \mathrm{derot}}}}{U_{\mathrm{obs}, \mathrm{derot}}}\right)^2 },
\end{equation} \label{ep}
where $\sigma_{Q_{\mathrm{obs}, \mathrm{derot}}}$ and $\sigma_{U_{\mathrm{obs}, \mathrm{derot}}}$ represent the root mean square of the neighboring noise region. In addition to the intrinsic PA profile, we also use the error-weighted average value $\overline{\mathrm{PA}_0}$ to estimate the population evolution. Whether this PA curve is flat or variable is determined by the reduced chi-square $\chi^2_{\nu}= \chi^2_{\mathrm{min}}/(N-n)$ of a constant PA fit following the conventional criterion \citep{2024ApJ...968...50P, 2025ApJ...988..175L}.
A varying PA type is adopted for PAs with $\chi^2_{\nu}>5$, whereas those with $\chi^2_{\nu}<5$ are categorized as flat.
$N$ is the number of PA points, and $n=1$ is the number of model parameters.
We caution that this classification is primarily intended as a descriptive characterization of the observed PA behavior, since the reduced chi-square depends on both intrinsic PA variability and measurement uncertainty, introducing a potential S/N dependence in the classification.
To avoid the noise-dominated PA points, we only retain the points with errors smaller than 5 degrees. In addition, we also test the influence of the uncertainty of RM with its posterior distribution from the $Q-U$ fitting method \citep{2019Sci...365.1013D}, and the induced uncertainty of PA is less than 1 degree, therefore we do not include it in Equation (\ref{ep}).

The time of arrival (TOA) of all bursts is converted to the barycentric framework at a certain reference frequency. The reference frequency is typically set at 1.5 GHz or at infinity frequency, which remains consistent within the same sample but may vary across different samples. We then merge the time and PA of different bursts into a single time-PA series in chronological order, which is ready for the PA-based period search.

\subsection{Lomb-Scargle Periodogram}
The PA is expected to reflect the magnetic configuration of the emission region.
For FRB models invoking rotating neutron stars or binary systems \citep{2025arXiv251207140W, 2025ApJ...994L..20Z}, rotation or orbital motion can periodically modulate the viewing geometry. Consequently, the line of sight (LOS) may sample different magnetic field orientations at different bursts, potentially producing a periodic variation in the PA time series.
To search for such periodicity, we apply the Lomb-Scargle periodogram (LSP) implemented in the Python package \texttt{ASTROPY} \citep{2022ApJ...935..167A}, which has been well-suited for unevenly sampled data and does not require a specific functional form of periods.

LSP is sensitive to the underlying periodic signal. When the trial frequency matched the real period, the phases of the periodicity aligned with the data, leading to coherent addition and producing a peak in the periodogram.
In the LSP, the power of a given frequency $f = 1/P$ can be calculated by
\begin{equation}
\begin{aligned}
 P_{\mathrm{LS}}(f) 
& =\frac{1}{2}\left\{\left(\sum_n g_n \cos \left(2 \pi f\left[t_n-\tau\right]\right)\right)^2 / \sum_n \cos ^2\left(2 \pi f\left[t_n-\tau\right]\right)\right. \\
& \left.+\left(\sum_n g_n \sin \left(2 \pi f\left[t_n-\tau\right]\right)\right)^2 / \sum_n \sin ^2\left(2 \pi f\left[t_n-\tau\right]\right)\right\},
\end{aligned}
\end{equation}
where $g_n$ is PA evaluated at the unevenly sampled times in this scenario.
$\tau$ is specified for each frequency to ensure time-shift invariance
\begin{equation}
\tau=\frac{1}{4 \pi f} \tan ^{-1}\left(\frac{\sum_n \sin \left(4 \pi f t_n\right)}{\sum_n \cos \left(4 \pi f t_n\right)}\right).
\end{equation}

The significance of the potential periodic signal is quantified using the false-alarm probability (FAP), which represents the probability that a signal without a periodic component would produce a peak power equal to or greater than the given power. 
This probability can be readily estimated through Monte Carlo simulations.
Specifically, we randomly shuffled the time series and conducted the LSP with the same frequency grid on this realization for 10000 realizations. For a given power, FAP can be estimated as the fraction of realizations in which the global maximum LSP power equals or exceeds it.

\begin{figure}[htbp]
    \centering
    \includegraphics[width=0.6\linewidth]{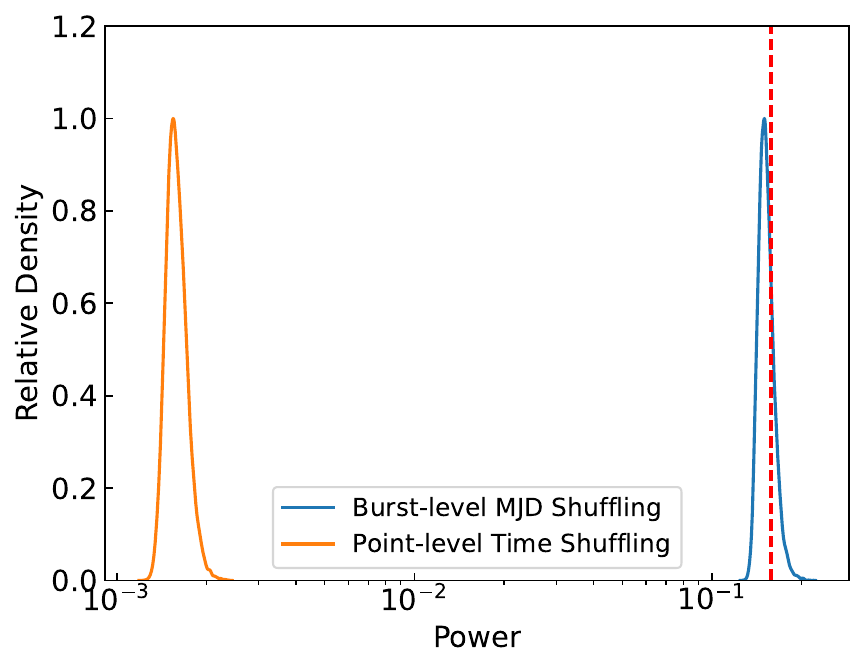}
    \caption{The kernel density estimation (KDE) of the Monte Carlo realizations of burst-level MJD shuffling and point-level time shuffling are shown in this figure. The peak of each curve has been normalized to unity. The red dotted line represents the maximum power of the first active episode of FRB 20201124A.}
    \label{fig: FAP}
\end{figure}

The time-PA series analyzed in this work exhibits a pronounced hierarchical sampling pattern, resembling a comb-in-comb structure. On short timescales, each burst consists of densely sampled PA measurements over its duration, forming a fine comb of regularly spaced data points. On longer timescales, the bursts themselves occur intermittently, with waiting time distributions showing characteristic clustering and gaps, effectively forming a second, much coarser comb. This nested sampling pattern introduces a complex observational window function that can imprint structured features in LSP. In Figure \ref{fig: FAP}, we show two methods of realizations generated using different time-randomization schemes. In the first approach, all individual time points are fully shuffled, while in the second approach, only the MJDs of bursts are randomized, preserving the internal temporal and polarization structure within each burst. The resulting LSPs demonstrate that the hierarchical comb-in-comb structure, together with the intra-burst structure, can introduce prominent aliasing features. Under these circumstances, the FAP derived from the fully shuffled realizations does not adequately capture the null hypothesis relevant to our problem. We therefore adopt the burst-MJD-shuffling scheme in all subsequent Monte Carlo simulations used to estimate the FAP.

The temporal coverage of our datasets is relatively long, spanning approximately 54 days for FRB 20201124A1, 4 days for FRB 20201124A2, 39 days for FRB 20220912A, and 489 days for FRB 20240114A. This extended time baseline enables us to probe periodic behavior on comparatively long timescales. Previous studies focusing on burst timescale periodicity have shown that no single, universal period can be identified, even though many individual bursts exhibit similar S-shaped PA swing \citep{2025ApJ...988..175L}. 
In this work, we concentrate on periodic behavior on timescales longer than the burst duration. For the full datasets, we therefore adopt a period search range of 10 ms to $10^{7}$ ms (about 2.778 hours). For individual observations during high-activity states, where the time span is about one hour, the period search range is restricted to 10 ms to $10^{6}$ ms (about 16.667 minutes).
The frequency range considered in the LSP is first divided into several logarithmic decades. Within each decade, $9 \times 10^6$ trial frequencies are generated with uniform linear spacing, corresponding to a trial-frequency grid spacing of $\Delta f = (f_{\mathrm{max}}-f_{\mathrm{min}})/(9 \times 10^6)$. $f_{\mathrm{max}}$, and $f_{\mathrm{min}}$ represent the maximum and minimum frequencies in this decade. For example, if we choose a decade of $[10^{-2}, 10^{-1})$ Hz, the spacing $\Delta f$ would be $10^{-8}$ Hz. The grid density was determined empirically by progressively increasing the number of trial frequencies until candidate peaks were adequately sampled. This strategy provides a sufficiently dense sampling of the frequency space, ensuring that potential periodic signals are not missed.

\subsection{Periodicity Search in Active Episode Datasets}
\begin{figure*}
\centering
\subfigure{\includegraphics[width=2.2in]{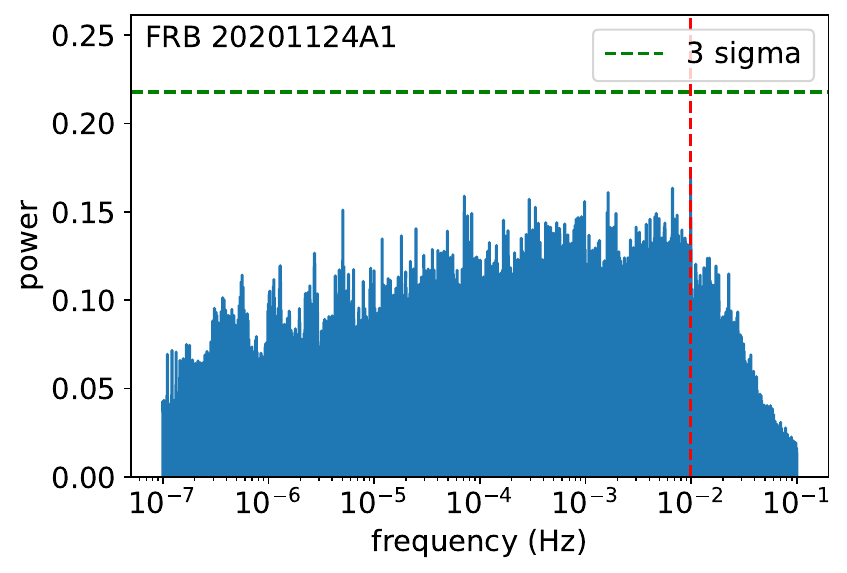}}
\subfigure{\includegraphics[width=2.2in]{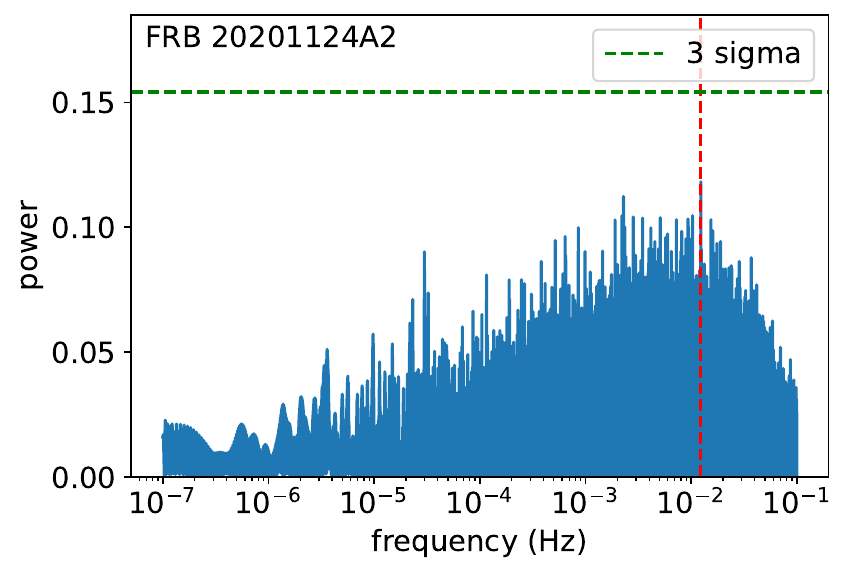}}
\subfigure{\includegraphics[width=2.2in]{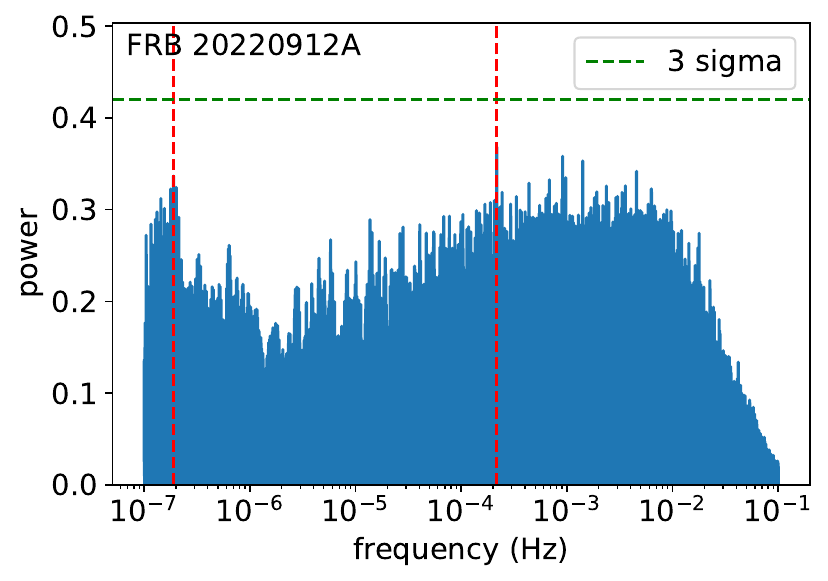}}
\subfigure{\includegraphics[width=2.2in]{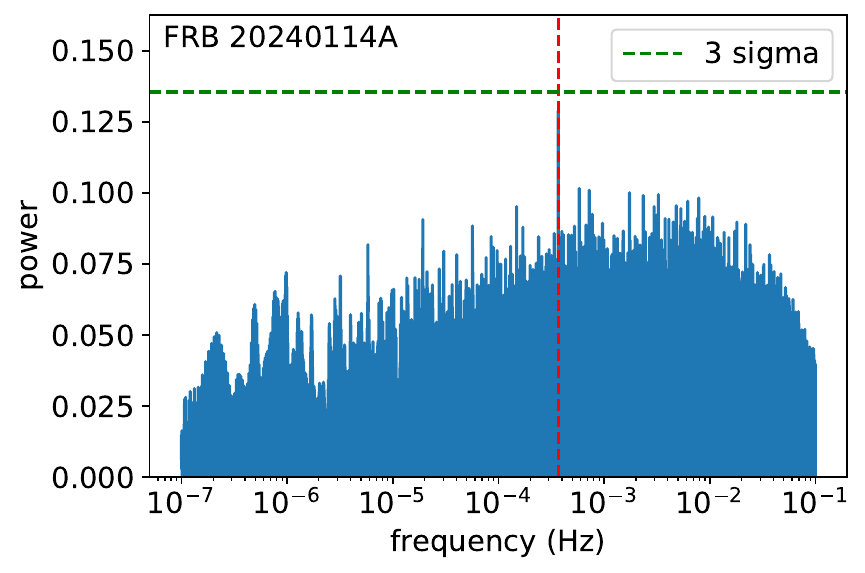}}
\caption{LSP search results using the time-PA series from four active episodes of FRB 20201124A, FRB 20220912A, and FRB 20240114A. The green dotted line represents the threshold corresponding to 3-sigma significance, and the red dotted vertical line shows the maximum power in our searching range.}
\label{fig: episode}
\end{figure*}

The LSP results for the four samples are shown in Figure \ref{fig: episode}. For all samples, no statistically significant periodic signal is detected, as the maximum LSP powers remain below the 3-$\sigma$ significance level.
In each sub-figure, the 3-$\sigma$ significance level is determined from 10000 Monte Carlo realizations constructed by randomly permuting the MJD of bursts.
For FRB 20201124A1, the highest LSP power occurs at a frequency of 0.009885041 Hz, with a FAP of 0.7782.
For FRB 20201124A2, the strongest peak is located at 0.01234387 Hz and has a FAP of 0.2503.
In both cases, the periodograms show a visually prominent peak at frequencies of order $10^{-2}$ Hz, although the associated statistical significances are low.
For FRB 20220912A, the maximum power appears at 0.0002165215 Hz, with a corresponding FAP of 0.2503. In addition, the periodogram exhibits a secondary peak at a much lower frequency of $1.893003 \times 10^{-7}$ Hz, with a FAP of 0.982, which indicates that it is likely attributed to the data structures.
As for FRB 20240114A, the maximum LSP power is found at a frequency of 0.0003690081 Hz, corresponding to a FAP of 0.0073. Although this represents the most significant candidate among the four samples, its significance is still below the 3-$\sigma$ threshold.
The corresponding FAP, which accounts for the full frequency search range, remains high, indicating that peaks of similar amplitudes can be easily generated by chance in realizations with the same observational sampling and statistical properties.
The LSP results obtained from single-day subsets give further support to the conclusion that this feature does not correspond to a genuine periodic signal.
Overall, the absence of statistically significant peaks suggests that the observed periodograms are consistent with effects induced by the temporal data structure rather than reflecting a physically meaningful spin or orbital period of the source.

\subsection{Periodicity Search in Single-day Observations}
\begin{figure*}
\centering
\subfigure{\includegraphics[width=2.2in]{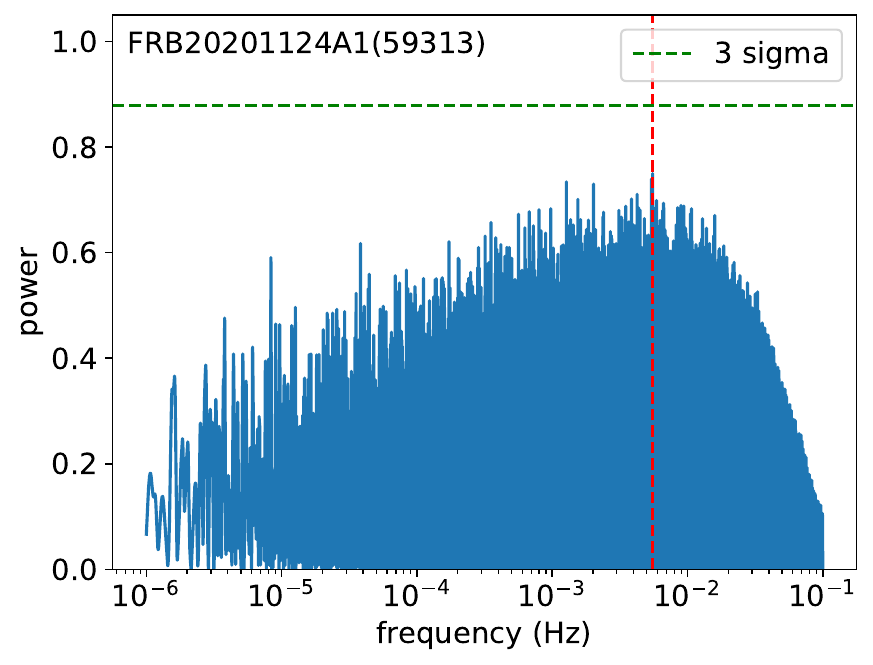}}
\subfigure{\includegraphics[width=2.2in]{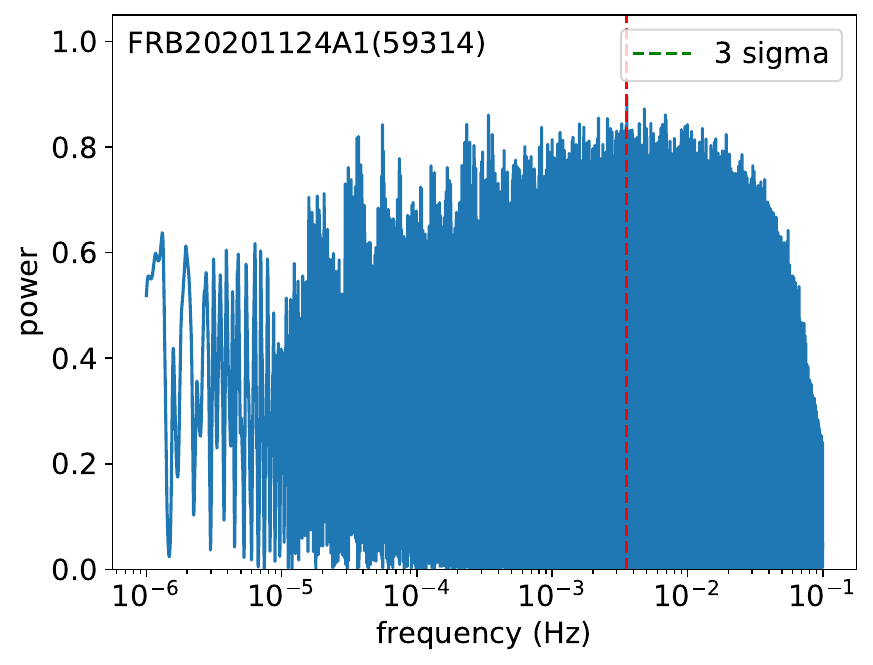}}
\subfigure{\includegraphics[width=2.2in]{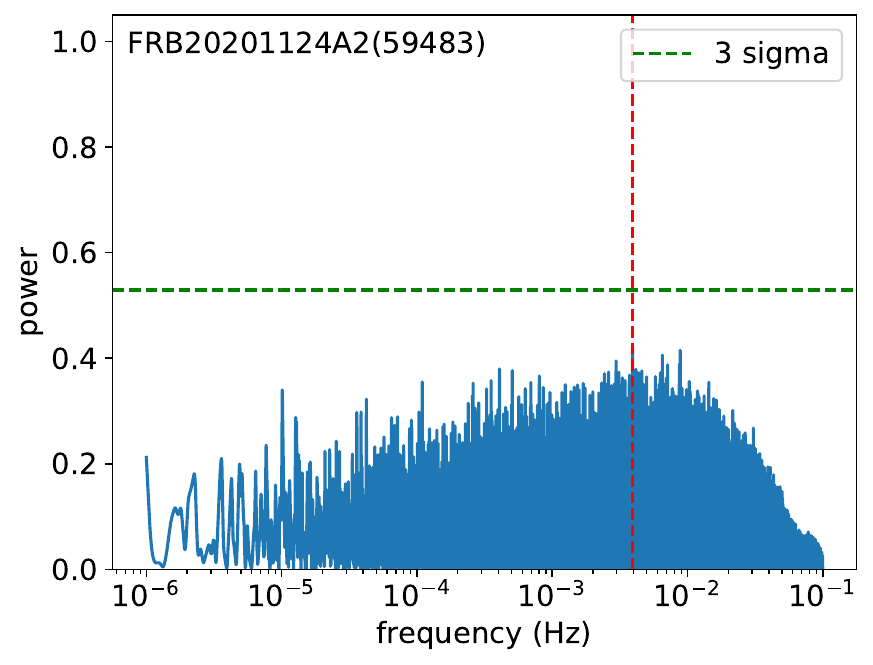}}
\subfigure{\includegraphics[width=2.2in]{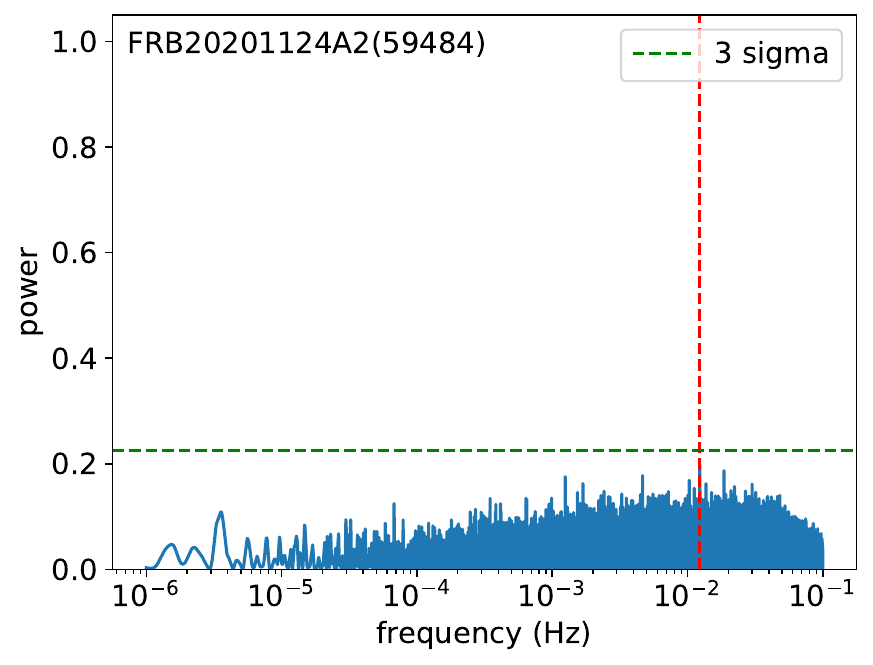}}
\subfigure{\includegraphics[width=2.2in]{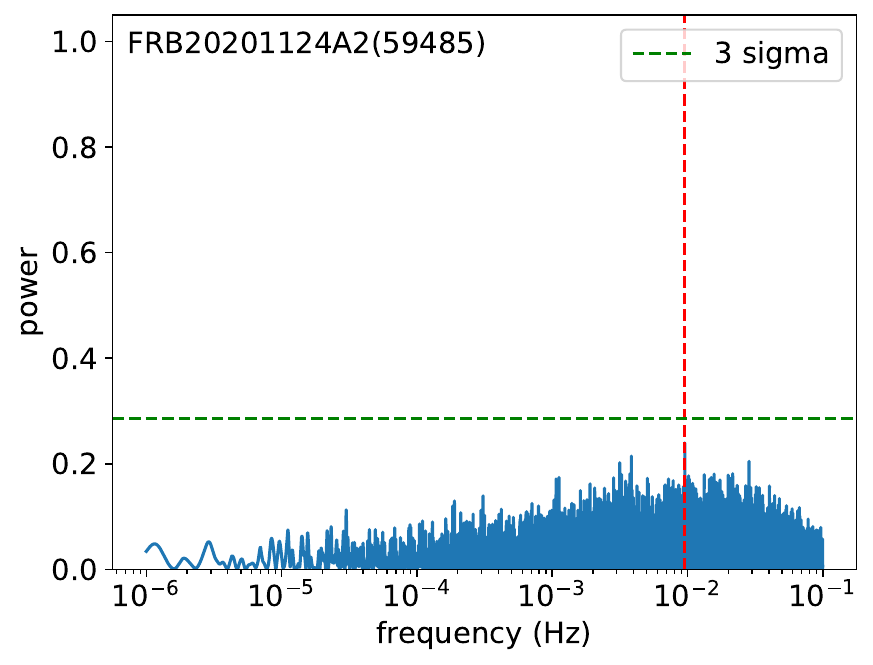}}
\subfigure{\includegraphics[width=2.2in]{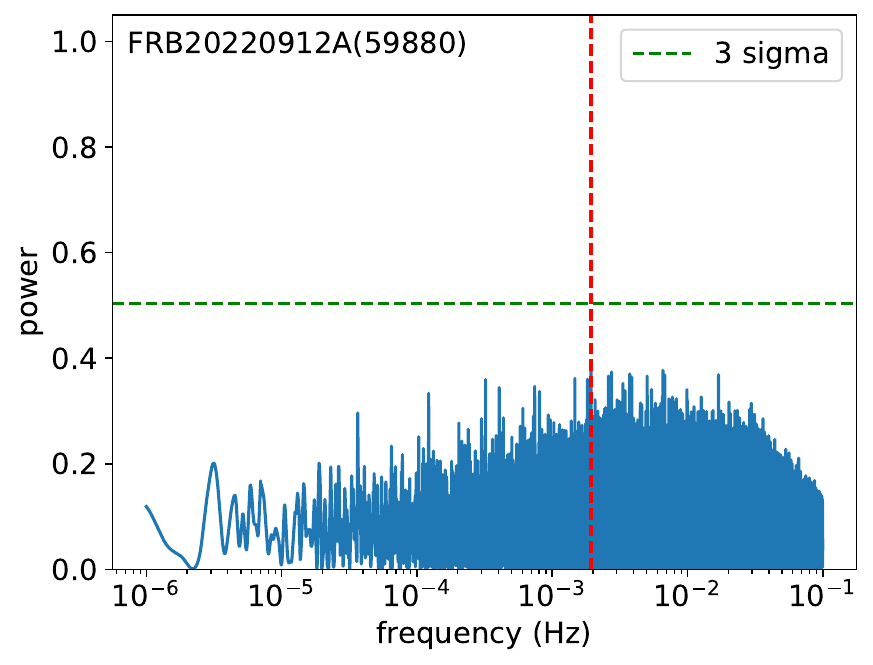}}
\subfigure{\includegraphics[width=2.2in]{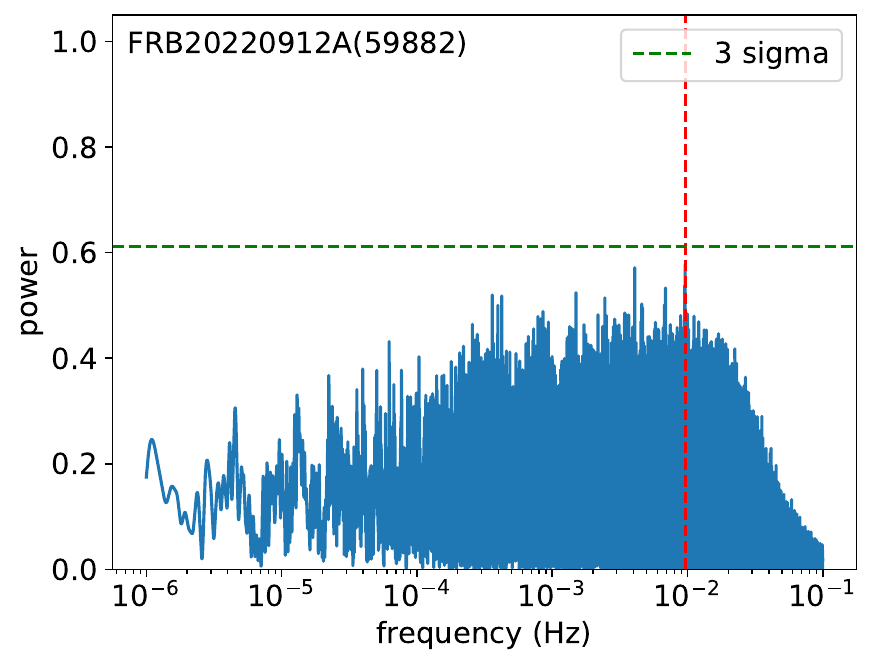}}
\subfigure{\includegraphics[width=2.2in]{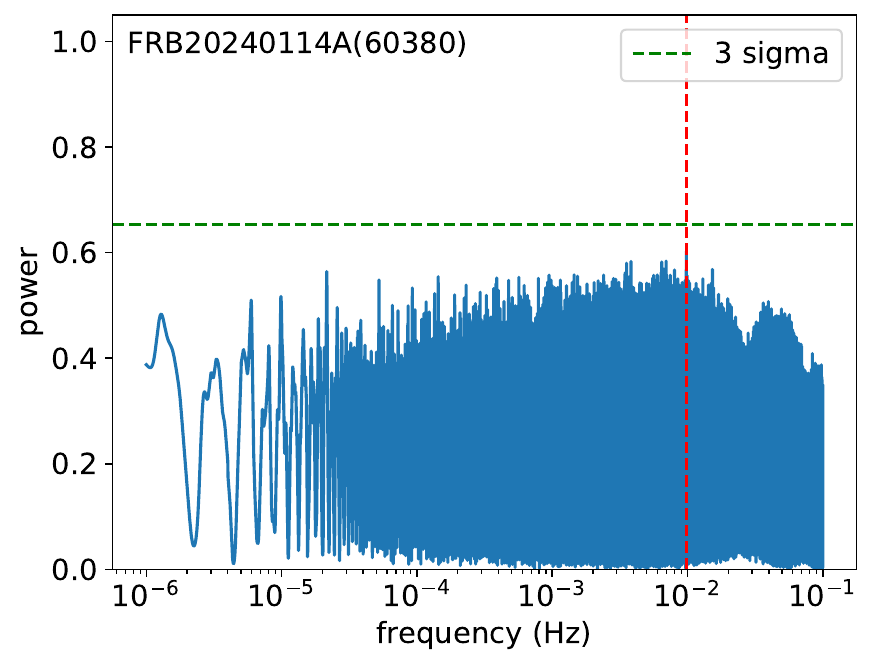}}
\subfigure{\includegraphics[width=2.2in]{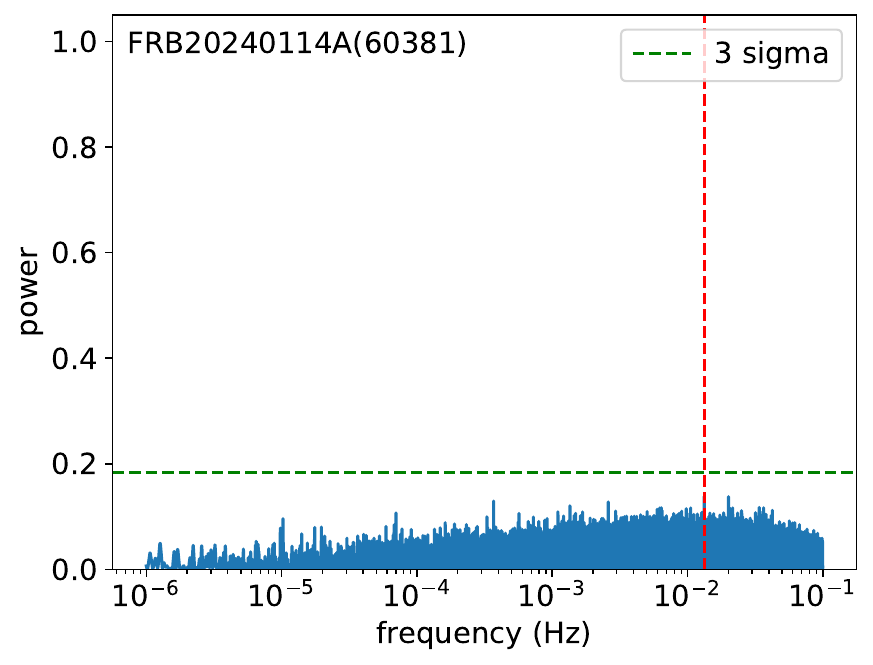}}
\caption{LSP search results using the time-PA series from several active days of FRB 20201124A and FRB 20220912A. The notation is the same as that in Figure \ref{fig: episode}.}
\label{fig: day}
\end{figure*}

We also perform LSP searches on individual observing days during which the sources exhibit relatively high activity, and the results are shown in Figure \ref{fig: day}.

For FRB 20201124A1, two active days (MJD 59313 and 59314) yield a peak at frequencies of 0.005524976 Hz and 0.004802489 Hz, respectively. These peak frequencies are lower than, but close to, the peak frequency identified in the FRB 20201124A1 sample.
FRB 20201124A2 shows similar results with the maximum LSP power found at 0.00392631 Hz (MJD 59483), 0.01234383 Hz (MJD 59484), and 0.009566257 Hz (MJD 59485). In particular, the peak frequency on MJD 59484 is very close to the peak frequency obtained from the FRB 20201124A2 sample, and the difference between these two frequencies is just 0.00000004 Hz, which also suggests that this peak is likely dominated by this single day rather than representing a global periodicity.
The peak frequencies of MJD 59880 and 59882 of FRB 20220912A are 0.00193869 Hz and 0.00961275 Hz, respectively.
Finally, the maximum single-day LSP power of FRB 20240114A appears at 0.0098277 Hz (60380) and 0.01330524 Hz (60381). Both these two peak frequencies are not consistent with the peak frequency 0.0003690081 Hz, which further supports that this candidate is not a real periodic signal.
In summary, while the peak frequencies derived from individual observations show noticeable scatter, only one single-day result, MJD 59484, reproduces a very similar peak frequency with the whole sample. Nevertheless, none of the single-day LSP power exceeds the 3-$\sigma$ significance threshold, and therefore the single-day searches alone do not provide statistically significant evidence for periodicity.

\section{Position Angle behaviors} \label{sec: PAb}

\subsection{Overview of PA Datasets}
\begin{figure}[htbp]
    \centering
    \includegraphics[width=0.6\linewidth]{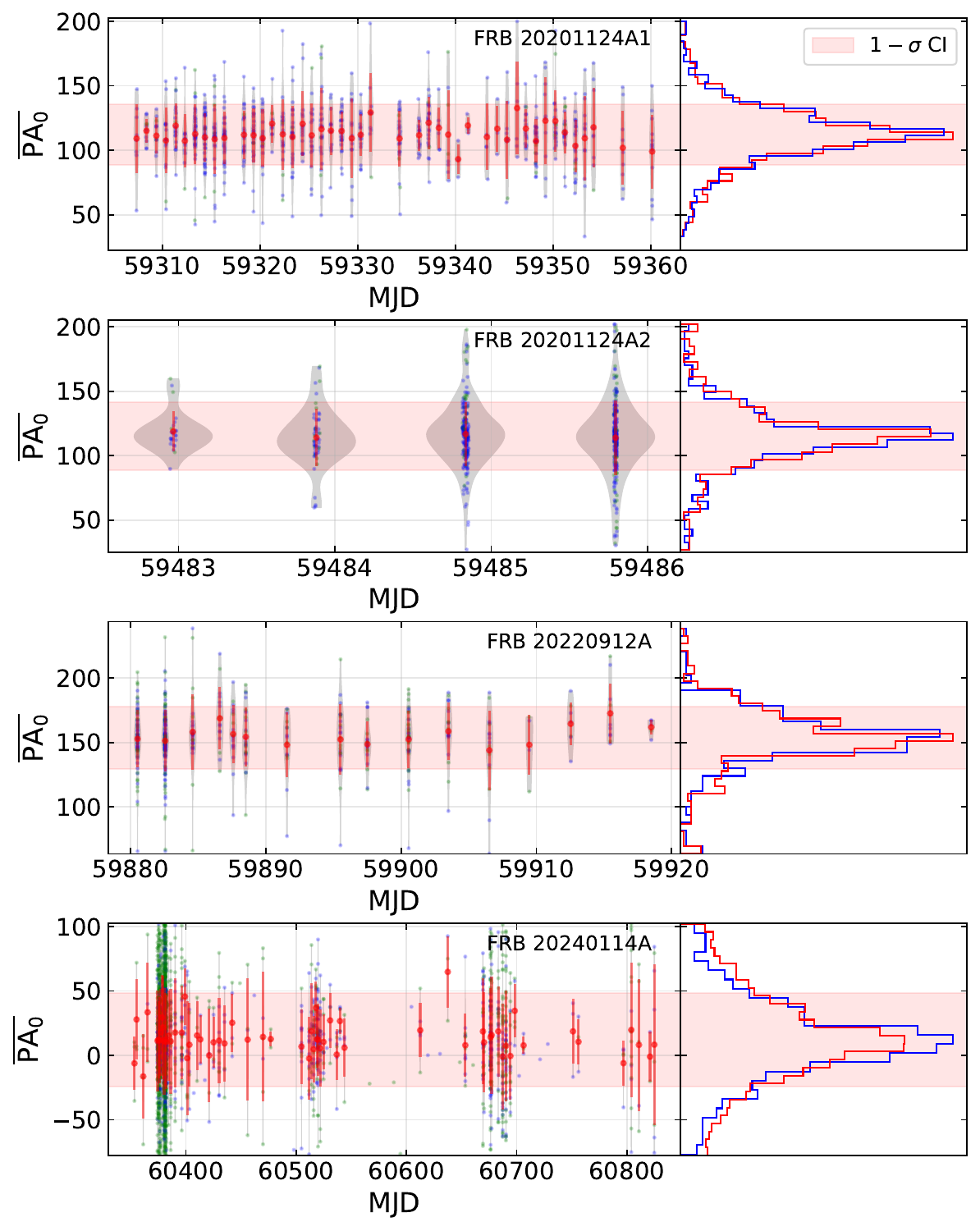}
    \caption{The $\overline{\mathrm{PA}_0}$ of bursts over MJD for FRB 20201124A1, FRB 20201124A2, FRB 20220912A, and FRB 20240114A, respectively. In the left panel of each sub-figure, the $\overline{\mathrm{PA}_0}$ of flat and variable PAs are marked in blue and green points, respectively. The grey KDE violins and the red points with the error bar denote the daily distribution of $\overline{\mathrm{PA}_0}$. In the right panel, we show the histogram of the $\overline{\mathrm{PA}_0}$ of all PAs and only flat PAs with red and blue colors, respectively. The light red region represents the 1-sigma confidence interval (CI), which is calculated using the whole sample. Note that the PAs have been shifted using their periodicity so that they are centered on the preferred angle.}
    \label{fig: PA0mjd}
\end{figure}

Given the recent interest in a nearly constant PA across bursts in FRB 20180916B \citep{2025A&A...702A.248B}, we also checked this burst-to-burst evolution in these samples. The $\overline{\mathrm{PA}_0}$ of each burst for the four datasets is shown in Figure \ref{fig: PA0mjd}. For all four samples, the PA varies significantly from burst to burst within a single observation.
We also investigated the PA differences of neighboring bursts and obtained consistent results.
Despite this burst-to-burst variability, their population properties exhibit a robust and well-defined structure. For each source, the PA values are tightly clustered around a preferred angle, forming a Gaussian-like distribution.
The Gaussian profile remains remarkably stable across all observations. Even when the variable PAs are included, the overall distribution remains unchanged. Both the position and width of the Gaussian profile are not universal but differ from one source to another.
In particular, FRB 20201124A1 and FRB 20201124A2 show statistically indistinguishable PA distributions, with consistent mean values and standard deviations, indicating no detectable evolution in the underlying PA population between these two active episodes.
More specifically, the PA distribution of FRB 20201124A1 follows a Gaussian distribution $\mathcal{N}(112.5, 23.5)$, and the PA distribution of FRB 20201124A2 follows a Gaussian distribution $\mathcal{N}(115.6, 26.3)$.
The PA distribution of FRB 20220912A can be fitted with a Gaussian distribution $\mathcal{N}(153.2,24.5)$.
The PA distribution of FRB 20240114A is modeled by a Gaussian distribution $\mathcal{N}(12.2,37.4)$.
All mean values and standard deviations are in units of degrees.
In conclusion, all four active episodes (three FRB sources) display remarkably similar phenomenology: large burst-to-burst PA variations coexisting with stable, Gaussian-like PA statistics over time.

\subsection{Structure Function of PA}
\begin{figure}[htbp]
    \centering
    \includegraphics[width=0.4\linewidth]{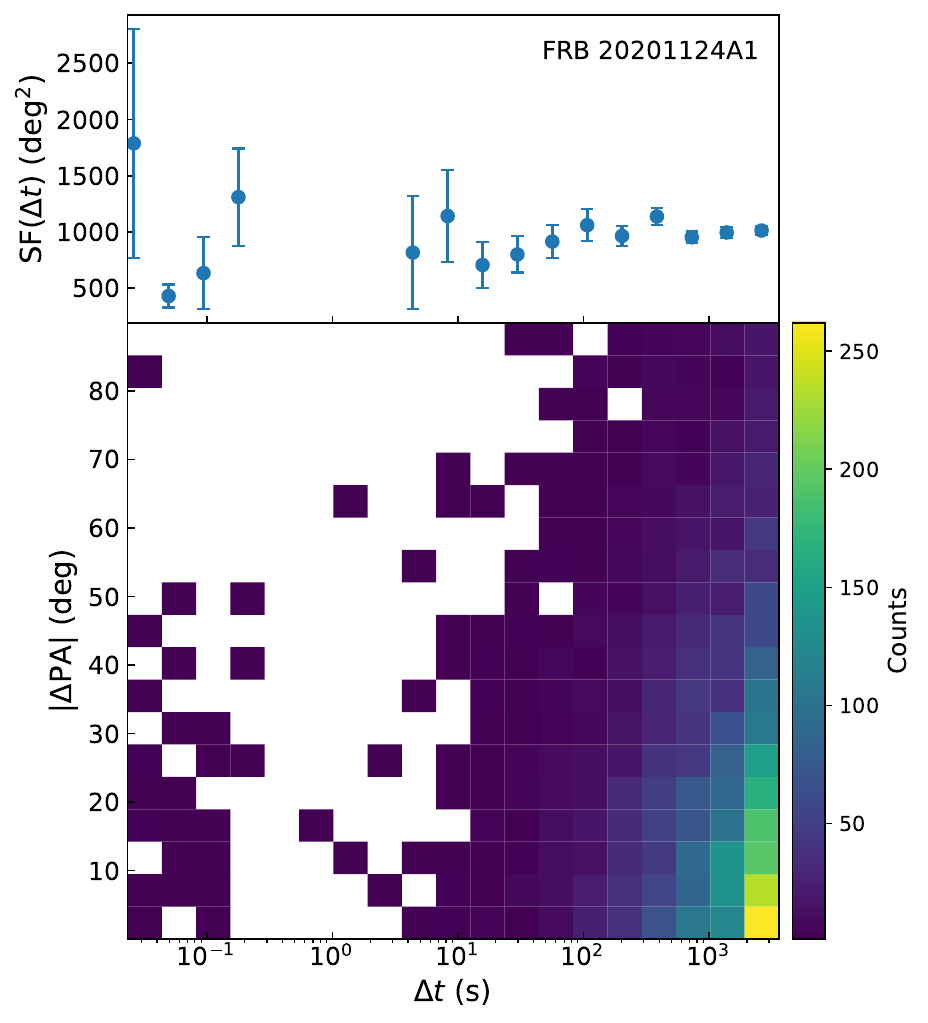}
    \includegraphics[width=0.4\linewidth]{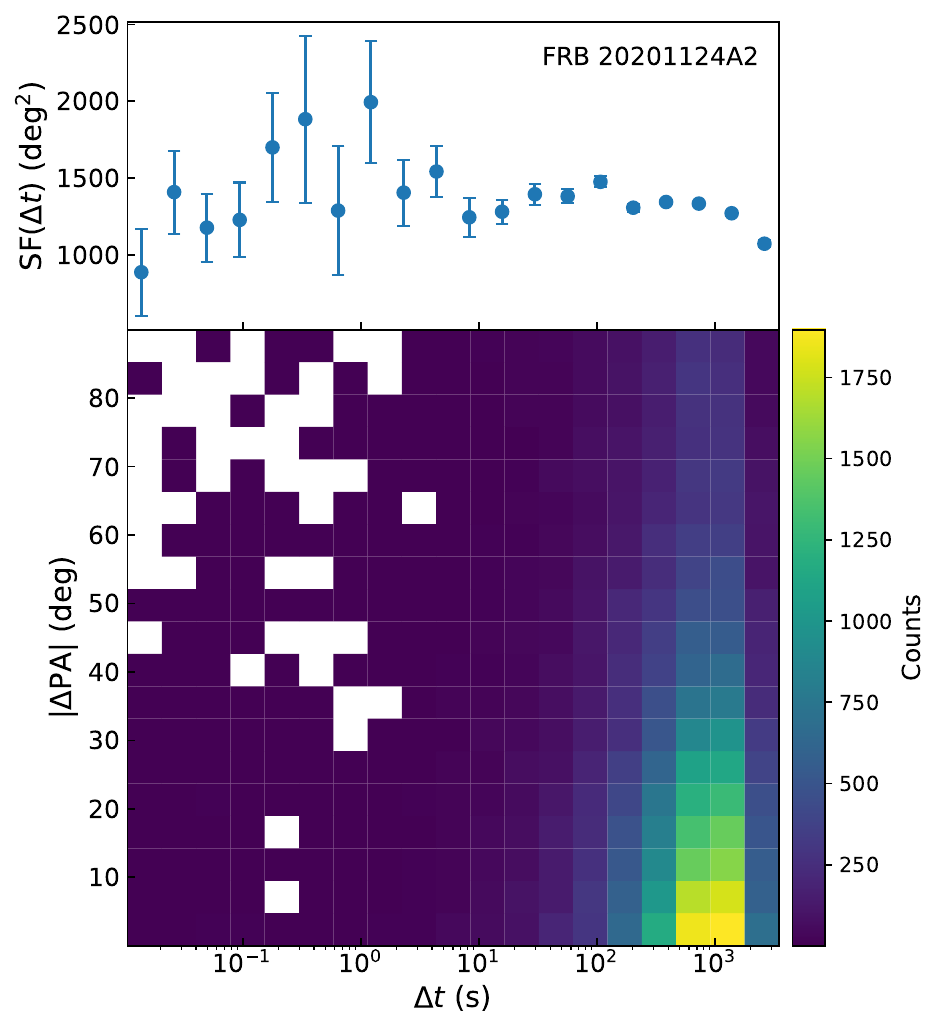}
    \includegraphics[width=0.4\linewidth]{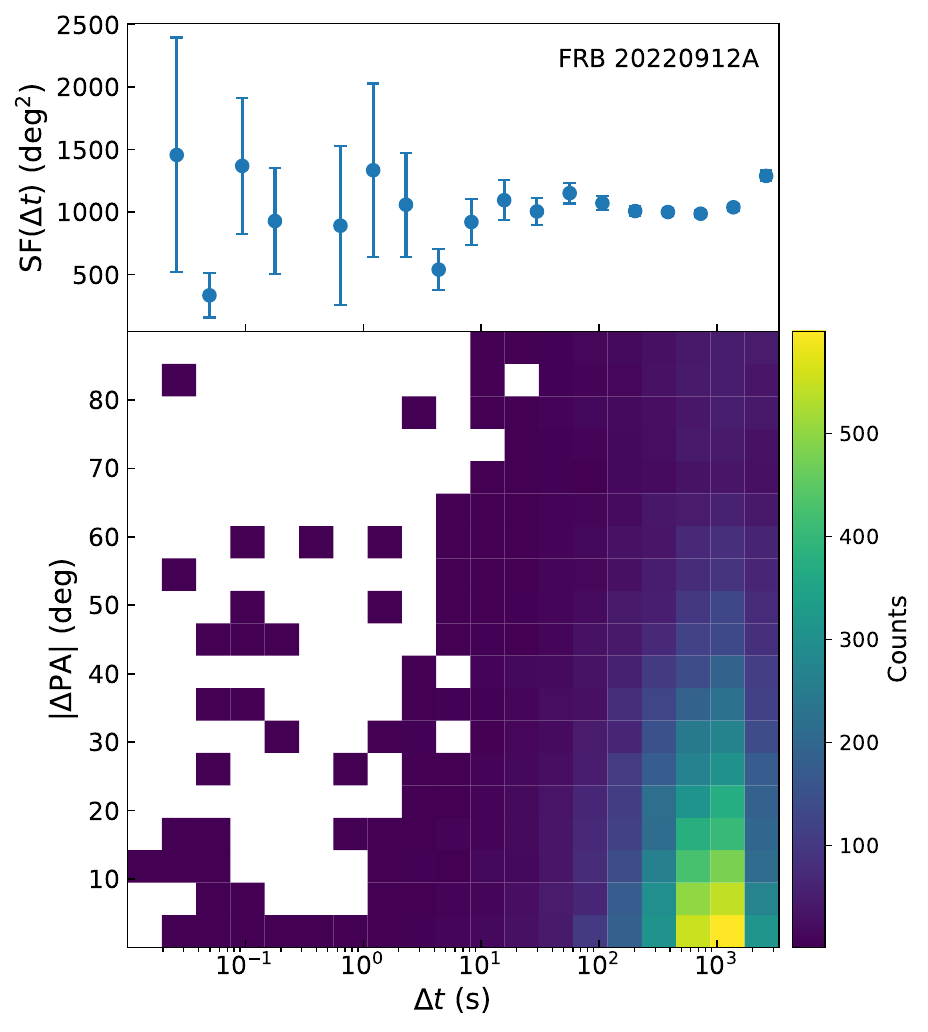}
    \includegraphics[width=0.4\linewidth]{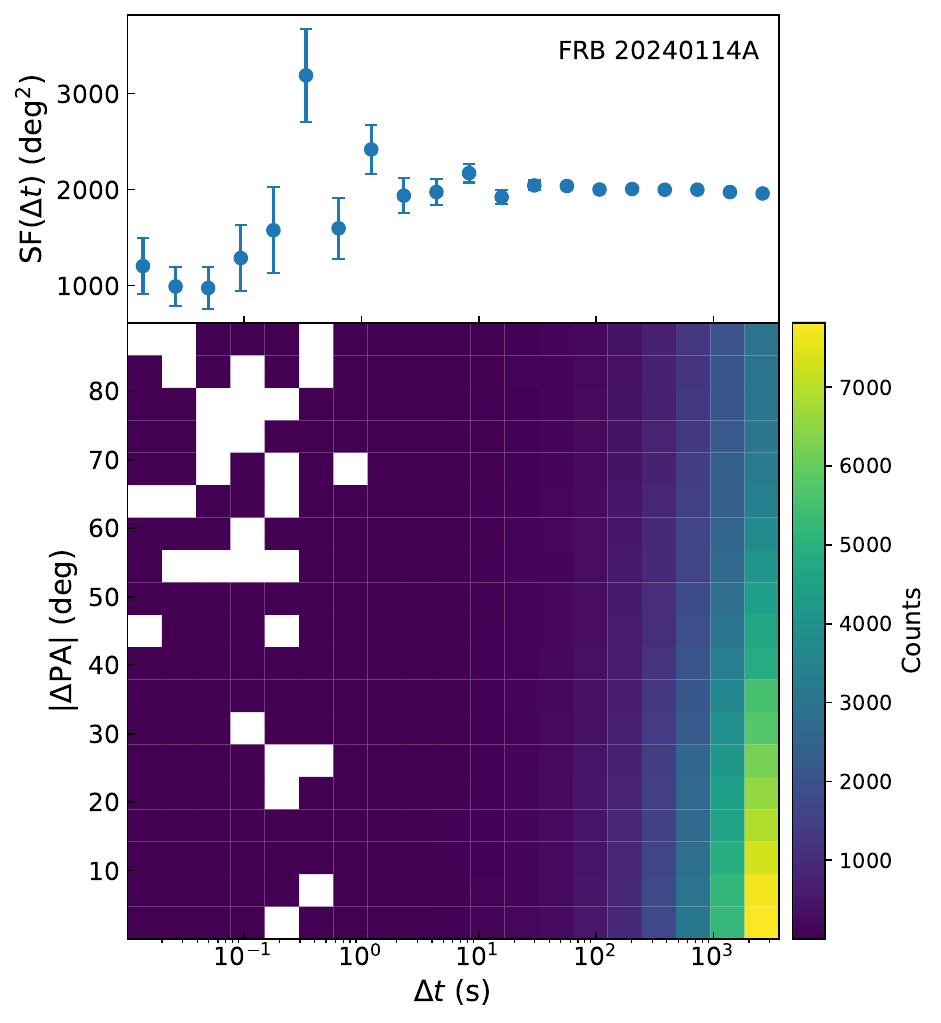}
    \caption{Two-dimensional distributions of all burst pairs (bottom panel) and the structure function $\mathrm{SF(\Delta t)}$ (upper panel) as a function of time lag. The time lag ranges from $10$ ms to 1 hour. The colorbar indicates the number of pairs in each bin.}
    \label{fig: SF}
\end{figure}

Given that the PA is thought to connect to the magnetic field line of the mission region, the random PA distribution suggests the magnetic configuration is dynamically evolving. Such a scenario may predict that bursts that are temporally closer have less PA difference, which can be directly tested using the structure function method. The SF method is a robust statistical tool to investigate the characteristic timescales of variability.
The 2-order SF is expressed as
\begin{equation}
    \mathrm{SF}(\Delta t) = \left< \left[ \mathrm{PA}(t+\Delta t) - \mathrm{PA}(t) \right]^2 \right>,
\end{equation}
where $\Delta t$ is the time lag.
In Figure \ref{fig: SF}, we show the 2-dimensional distributions of all observed pairs and the corresponding SFs for each dataset.
The results suggest that, over the considered time lag scale from $10^{-2}$ s to $3600$ s (1 hour), the SF is substantial, exceeding 1000 deg$^2$.
For FRB 20201124A1, FRB 20201124A2, and FRB 20220912A, the SF values remain constant across the entire range of time lags. Conversely, the SF of FRB 20240114A exhibits a distinct evolutionary trend: at short time lags, the SF is approximately half of that observed at long time lags, while remaining remarkably large even at the shortest lags.
These collective behaviors indicate that the underlying process has a randomized timescale shorter than 10 ms.

\subsection{Influence of the Rotation Measure Evolution}
The RM of FRB 20201124A shows a long-term evolutionary pattern with a potential period of 26.24 days \citep{2025arXiv250506006X}, ranging from $-800$ to $-400\,\mathrm{rad \text{ } m^{-2}}$ \citep{2022Natur.609..685X, 2022RAA....22l4001Z, 2022RAA....22l4003J}. The data used in this paper only covers two active episodes, while the PA distribution remains consistent from beginning to end. In early observations of repeaters, burst-to-burst PA variability was noticed \citep{2021MNRAS.508.5354H, manaswini2026polarizationpositionanglejumps}, while the limited number of bursts prevented further analysis.
The RM of FRB 20220912A is almost zero and remains stable over the FAST observation period \citep{2023ApJ...955..142Z}.
In the observation of FRB 20240114A, the RM ascended to $\sim$400 $\mathrm{rad \text{ } m^{-2}}$ initially, but declined to $\sim$200 $\mathrm{rad \text{ } m^{-2}}$ by the end of the observation period \citep{Wang:2026ssk}.
Similar PA behaviors among diverse local environments of FRB sources suggest that PA is an intrinsic property related to the emission mechanism and is affected very little by the propagation effects.

\subsection{Comparison with Other Observations}
For FRB 20240114A, \cite{2026arXiv260216409U} collects a sample of 5526 bursts using the ultrawideband receiving system of the Parkes radio telescope, whose PA measurements also show time-independent behavior, but with a bimodal PA distribution. Given the overlapping observational periods of FAST and Parkes, the observed discrepancy cannot be attributed to temporal evolution, but instead points to a chromatic behavior of the PA or the existence of a secondary magnetic axis.

Although we observe significant burst-to-burst PA variations in FRB 20201124A, FRB 20220912A, and FRB 20240114A, such behavior seems not to be universal among repeating FRBs. In contrast, sources with longer timescale periods, such as FRB 20121102A and FRB 20180916B, exhibit flat PA profiles within individual bursts and remain nearly consistent across different bursts in a single observation \citep{2018Natur.553..182M, 2018ApJ...863....2G, 2025A&A...702A.248B}, suggesting a rather stable magnetic configuration.
However, observations of FRB 20180916B reveal noticeable PA variations between different observing epochs, with the PA appearing to cluster around a preferred angle \citep{2025A&A...702A.248B}, a behavior qualitatively similar to that seen in our samples. This may suggest that periodic and aperiodic FRBs may not exhibit fundamentally different polarization behaviors.
The polarimetric properties of more periodic FRBs would provide important information on whether there is a correlation between the long-term cycle and the activity level of the magnetosphere.

\section{The role of a non-stable magnetosphere} \label{sec: sim}
\subsection{Simulation of the Dynamically Evolving Magnetosphere}
Previous works have proposed that the geometric configuration, particularly the relative alignment of the magnetic and spin axes, may explain the observed difference between repeating and one-off FRBs \citep{2025ApJ...982...45B}. Such interpretations generally assume a stable magnetosphere.
However, the observed PA behavior of FRBs exhibits diverse and non-periodic evolution, rather than a regular evolution predicted by the simple RVM. This discrepancy suggests that the magnetic axis responsible for bursts may not remain stable.
If the magnetosphere evolves dynamically, the effective magnetic axis can vary with time. Variations on millisecond timescales would dominate the intra-burst PA swing, while longer timescales may lead to a burst-to-burst distribution of PA.
To account for this long-term behavior, we extend the standard RVM by allowing the magnetic axis to evolve between bursts, resulting in a dynamical RVM framework.

The perturbation to the original magnetic axis $\mathbf{m}$ is assumed to be uniform in all directions within the vertical plane, and the angle change of the magnetic axis can decrease with the distance according to a probability distribution.
In order to facilitate the normalization of the probability distribution, we use the three-dimensional von Mises-Fisher distribution to model the effective magnetic axes in the unit sphere, whose density function is given by
\begin{equation}
    f(\mathbf{m}_{\mathrm{eff}})=\frac{\kappa^{1/2}}{(2 \pi)^{3/2} I_{1/2}(\kappa)} \exp (\kappa \mathbf{m} \cdot \mathbf{m}_{\mathrm{eff}})
\end{equation}
where $\mathbf{m}$ is the magnetic axis without perturbation, and $\mathbf{m}_{\mathrm{eff}}$ is the effective magnetic axis. $I_{1/2}$ is the modified Bessel function of the first kind. $\kappa$ is the concentration parameter, with larger $\kappa$ indicating a narrower distribution.
In the small-angle limit (i.e., when the effective magnetic axes are tightly concentrated around the magnetic axis), the von Mises-Fisher distribution can be approximated as a Gaussian distribution on the tangent plane. In this region, the concentration parameter satisfies $\kappa \approx 1/\sigma^2$, where $\sigma$ represents the standard deviation of the locally approximating isotropic normal distribution.
In the following, we will use $\sigma$ instead of $\kappa$, as it carries a more intuitive physical meaning. However, it should be noted that this correspondence breaks down when the small-angle approximation is no longer valid.
In the simulations, the effective magnetic axis is sampled using \texttt{scipy.stats.vonmises\_fisher} from the \texttt{SciPy} package.

\begin{figure*}
\centering
\subfigure{\includegraphics[width=2.3in]{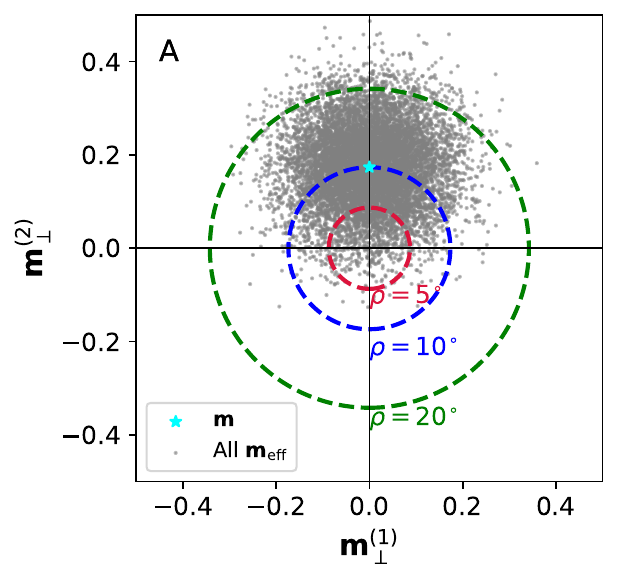}}
\subfigure{\includegraphics[width=2.3in]{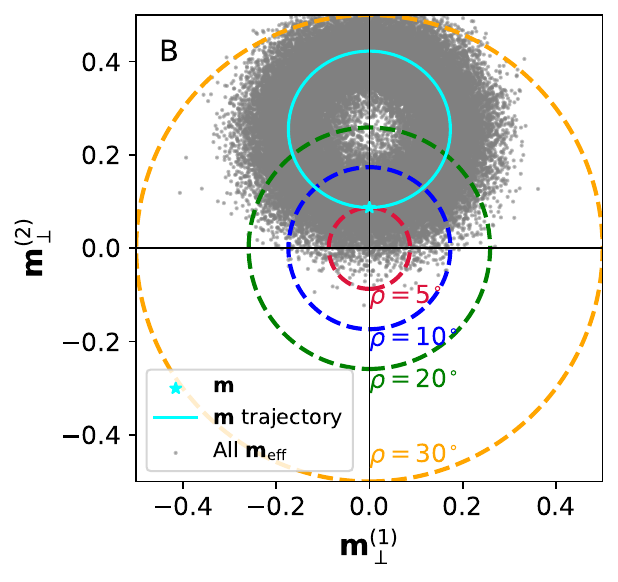}}
\subfigure{\includegraphics[width=2.3in]{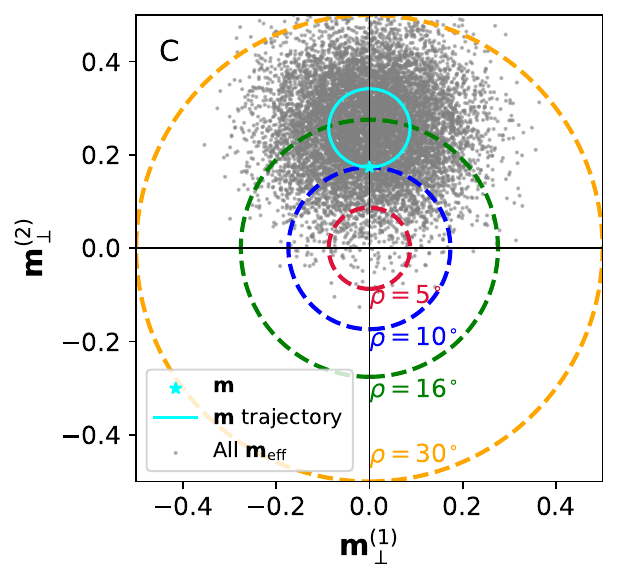}}
\caption{Distribution of the effective magnetic axes $\mathbf{m}_{\mathrm{eff}}$ on the sky plane for these three scenarios. The grey dots denote the projection of the effective magnetic axes, and the cyan star marks the projection of the mean magnetic axis $\mathbf{m}$ at the fiducial plane. The dashed circles represent the visible regions corresponding to different half-opening angles $\rho$. The cyan circles in panels B and C represent the trajectory of $\mathbf{m}$. $\mathbf{A}$: Non-rotating case, where $\mathbf{m}$ is fixed and $\mathbf{m}_{\mathrm{eff}}$ is generated from the 2-D von Mises-Fisher distribution. The parameter settings of the panel A are $\{\alpha, \beta, \sigma \} = \{ 20^{\circ}, 10^{\circ}, 5^{\circ} \}$. $\mathbf{B}$: The rotating case with small perturbations and the corresponding parameter settings of panel A are $\{\alpha, \beta, \sigma \} = \{ 10^{\circ}, 5^{\circ}, 3^{\circ} \}$. $\mathbf{C}$: The rotating case with large perturbations and the corresponding parameter settings of panel A are $\{\alpha, \beta, \sigma \} = \{ 5^{\circ}, 10^{\circ}, 5^{\circ} \}$.}
\label{fig: sim}
\end{figure*}

Since the distribution of $\mathbf{m}_{\mathrm{eff}}$ is isotropic, the PA evolution of the mean magnetic axis still follows the prediction of RVM, which can be described by
\begin{equation}
\tan \left(\mathrm{PA} - \mathrm{PA}_0 \right)=\frac{\sin \alpha \sin \left( \phi - \phi_0 \right)}{\sin\left( \alpha + \beta \right) \cos (\alpha)-\cos \left( \alpha + \beta \right) \sin \alpha \cos \left( \phi - \phi_0 \right)},
\label{eq:tanPA}
\end{equation}
where the inclination angle $\alpha$ represents the angle between the spin axis and the mean magnetic axis, the impact angle $\beta$ represents the minimum angle between LOS and the mean magnetic axis, and $\phi$ denotes the rotation phase.
$\mathrm{PA}_0$ and $\phi_0$ represent the reference PA and phase, respectively, and are both set to zero.
We assume that the rotation is around the z-axis and the fiducial plane is the same as the x-z plane.
Actually, when the perturbations are included, our dynamic RVM only adds one parameter, $\sigma$. In the following simulation, we consider three scenarios: A) a non-rotating magnetic axis, B) a rotating magnetic axis with small perturbations ($\sigma \ll \alpha$), and C) a rotating magnetic axis with large perturbations ($\sigma \geq \alpha$). Scenario A is the simplest case, where a fixed mean magnetic axis lies within the x-z plane.
In scenario B, the characteristic width $\sigma$ is smaller than the inclination angle $\alpha$, so that $\mathbf{m}_{\mathrm{eff}}$ fluctuates around $\mathbf{m}$ within a small angular region and does not cross the rotation axis. In scenario C, $\sigma$ becomes comparable to or larger than $\alpha$, allowing the magnetic axis to explore a much broader solid angle. In Figure \ref{fig: sim}, we show the projections of the effective magnetic axes on the sky plane for these three scenarios.

\subsection{PA Distribution of the Dynamical RVM}
\begin{figure*}
\centering
\subfigure{\includegraphics[width=2.3in]{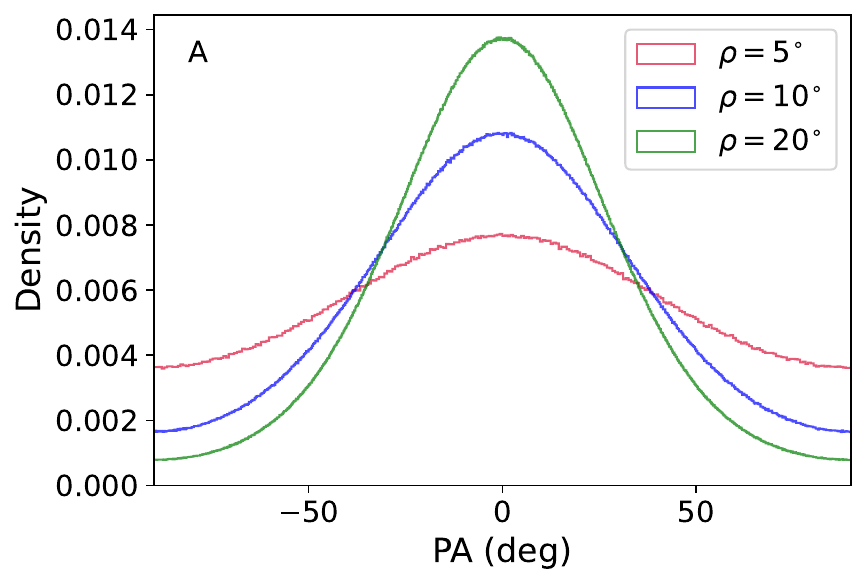}}
\subfigure{\includegraphics[width=2.3in]{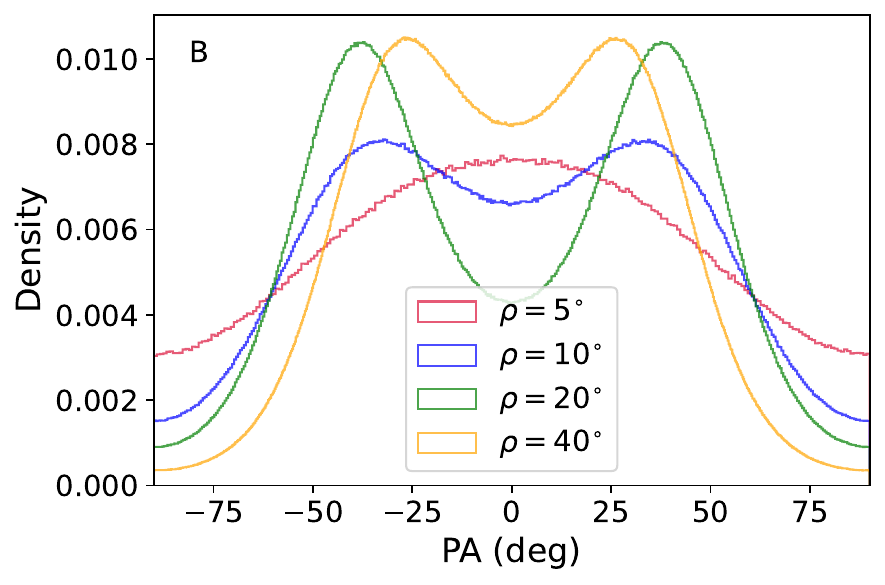}}
\subfigure{\includegraphics[width=2.3in]{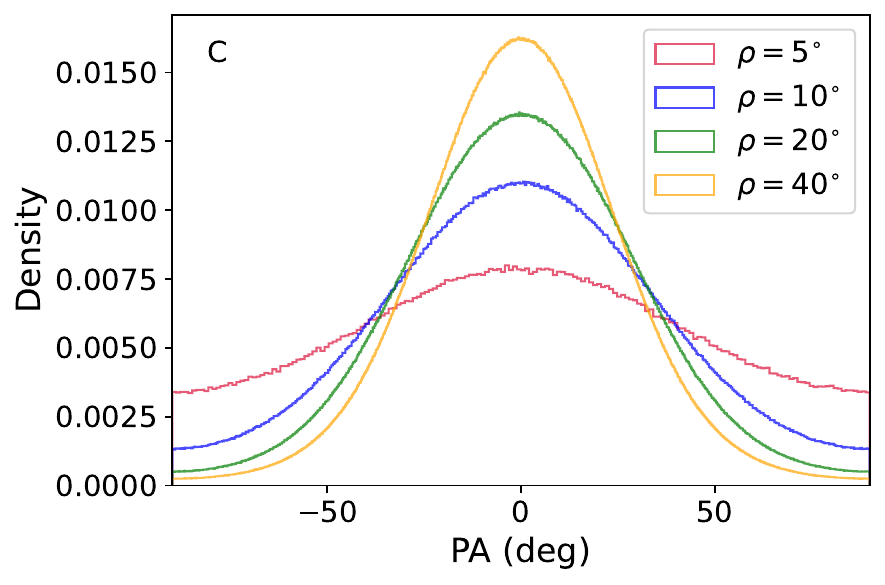}}
\caption{PA distributions for different half-opening angles in scenarios A, B, and C. Within each scenario, curves with the same color correspond to the same value of $\rho$ in Figure \ref{fig: PA0}.}
\label{fig: PA0}
\end{figure*}

\begin{figure*}
\centering
\subfigure{\includegraphics[width=3in]{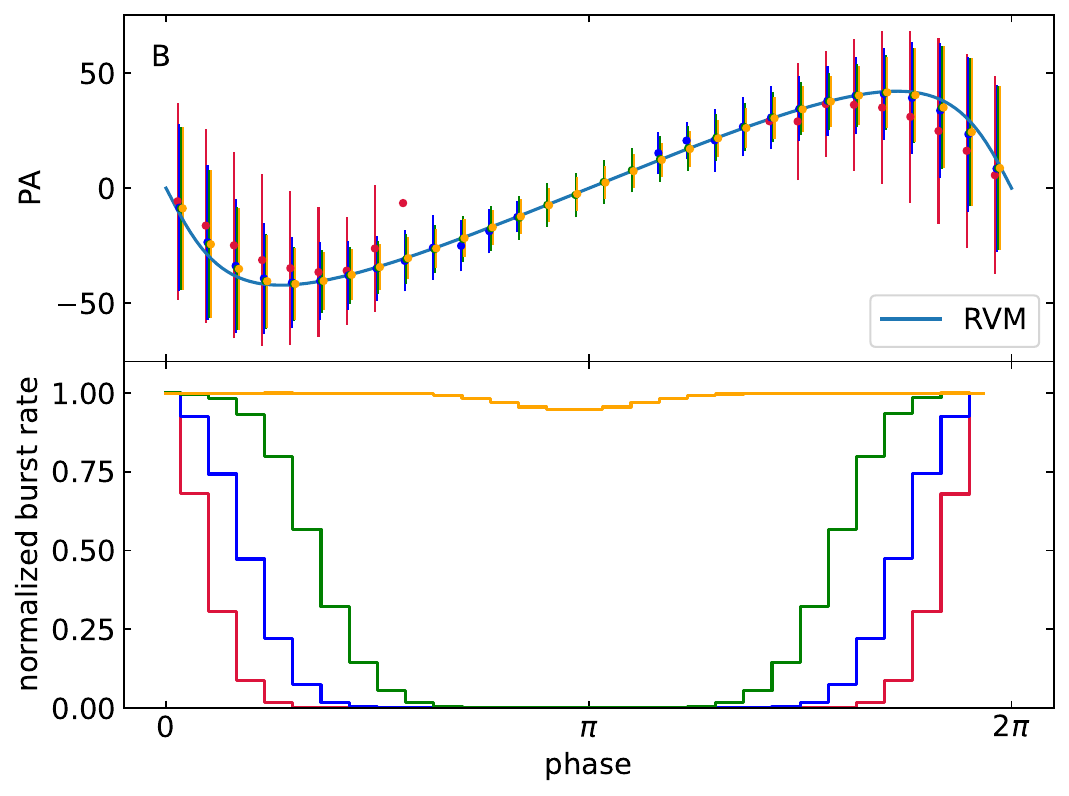}}
\subfigure{\includegraphics[width=3in]{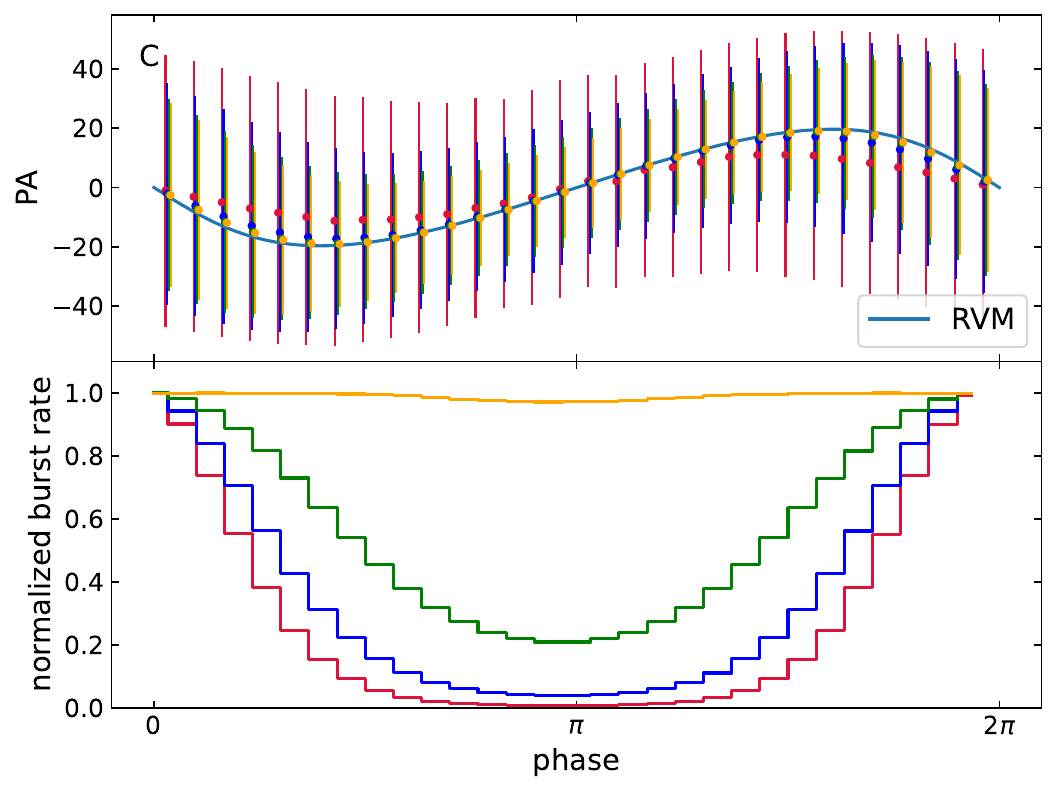}}
\caption{The normalized burst rate and the PA distribution of 30 phase bins in scenarios B and C. Within each scenario, curves and points with the same color correspond to the same value of $\rho$ in Figure \ref{fig: PA0}.}
\label{fig: phase}
\end{figure*}

We show the PA distributions of these three scenarios in Figure \ref{fig: PA0}.
In all panels, we consider different half-opening angles $\rho$, ranging from values that cover only a small portion to those that encompass nearly the entire effective magnetic axes.
In scenario A, the PA distributions remain unimodal for all half-opening angles, and the distribution becomes narrower with increasing $\rho$.
A similar situation also occurs in scenario C, while the PA distributions of scenario B show a symmetrical bimodal structure when the half-opening angle $\rho$ is larger than the impact angle $\beta$.
The noticeable difference arises from the fact that the distribution of PA is dependent on the rotation phase. However, when the perturbation becomes sufficiently strong, the intrinsic distribution of the effective magnetic axis becomes broad, such that the cone traced by the mean magnetic axis remains largely confined within the central region of the distribution. In this case, the phase-dependent variation in the mean PA is significantly suppressed.
To better characterize this behavior, we therefore divide these simulated bursts into 30 phase bins, and the corresponding normalized burst rate and the PA distributions are shown in Figure \ref{fig: phase}. In the small perturbation scenario B, bursts almost disappear when the magnetic axis turns to the back side ($\phi \sim \pi$), and the normalized burst rate would show a regular on-off phase. 
The normalized burst rate exhibits significant periodic modulation due to the rotation of $\mathbf{m}$. When $\rho > \beta+2\alpha$, the modulation is barely discernible. However, the coevolution of the mean and standard deviation of the PA distribution for each phase bin persists under this condition.
The PA distribution also exhibits a phase evolution, characterized by an RVM-like trend in its mean and a suppressed standard deviation on the back side, which provides a direct observable feature in the data.

\subsection{Application on Our Datasets}
Searching for the coevolution behavior of the mean and standard deviation of the PA distribution provides a novel approach to identify the missing period from the PA population of repeating FRBs.
The essential task in period searching is to construct a diagnostic quantity that responds sensitively to the true period. In this method, we use the reduced chi-square $\chi^2_{\nu}$ of the mean and standard deviation of the PA distribution within each phase bin to achieve this.
When the correct period is used, the binned mean PA would show RVM-like trend and the binned standard deviation also decreases in the back side as simulated in Figure \ref{fig: phase}. 
In contrast, when an incorrect trial period is adopted, the folded PA distribution is effectively randomized in phase, leading to nearly flat curves for both the mean and the standard deviation.
This method relies on population-level statistics and therefore requires a sufficiently large sample of bursts. Consequently, we use all $\overline{\mathrm{PA}_0}$ values in the samples rather than restricting the analysis to the flat PA sub-samples. This approach may become less effective for PA curves with large PA swings, but it remains applicable to the majority of cases considered here.

Our procedure is implemented as follows. For a given trial period $P_i$, we fold the MJDs of all bursts into phase and divide the phase interval into 18 equally spaced bins. Within each bin, we compute the mean $\mu_j$ and standard deviation $\sigma_j$ of $\overline{\mathrm{PA}_0}$, together with their uncertainties, and calculate the corresponding reduced chi-square values, $\chi^2_{\nu,\mu}$ and $\chi^2_{\nu,\sigma}$, which are defined as
\begin{equation}
\begin{aligned}
&\chi^2_{\nu,\mu} = \frac{1}{N_{\mathrm{bins}}-1} \sum_{j=0}^{N_{\mathrm{bins}}} \frac{( \overline{\mu} - \mu_j )^2}{ \Delta \mu_j^2},\\
&\chi^2_{\nu,\sigma} = \frac{1}{N_{\mathrm{bins}}-1} \sum_{j=0}^{N_{\mathrm{bins}}} \frac{ (\overline{\sigma} - \sigma_j)^2 }{\Delta \sigma_j^2},
\end{aligned}
\end{equation}
where $N_{\mathrm{bins}}-1$ is the degree of freedom. The uncertainties $\Delta \mu_j$ and $\Delta \sigma_j$ are estimated from $\sigma_j/\sqrt{N}$ and $\sigma_j/\sqrt{2N}$, respectively.
$\overline{\mu}$ and $\overline{\sigma}$ are the best-fit mean and standard deviation determined by minimizing the chi-square.
This process is repeated for all trial periods in the range $[1, 600]$ s, sampled with a step size of $10^{-6}$ s. The correct period is expected to produce a simultaneous enhancement in both $\chi^2_{\nu,\mu}$ and $\chi^2_{\nu,\sigma}$ relative to the noise level.

\begin{figure*}
\centering
\subfigure{\includegraphics[width=3in]{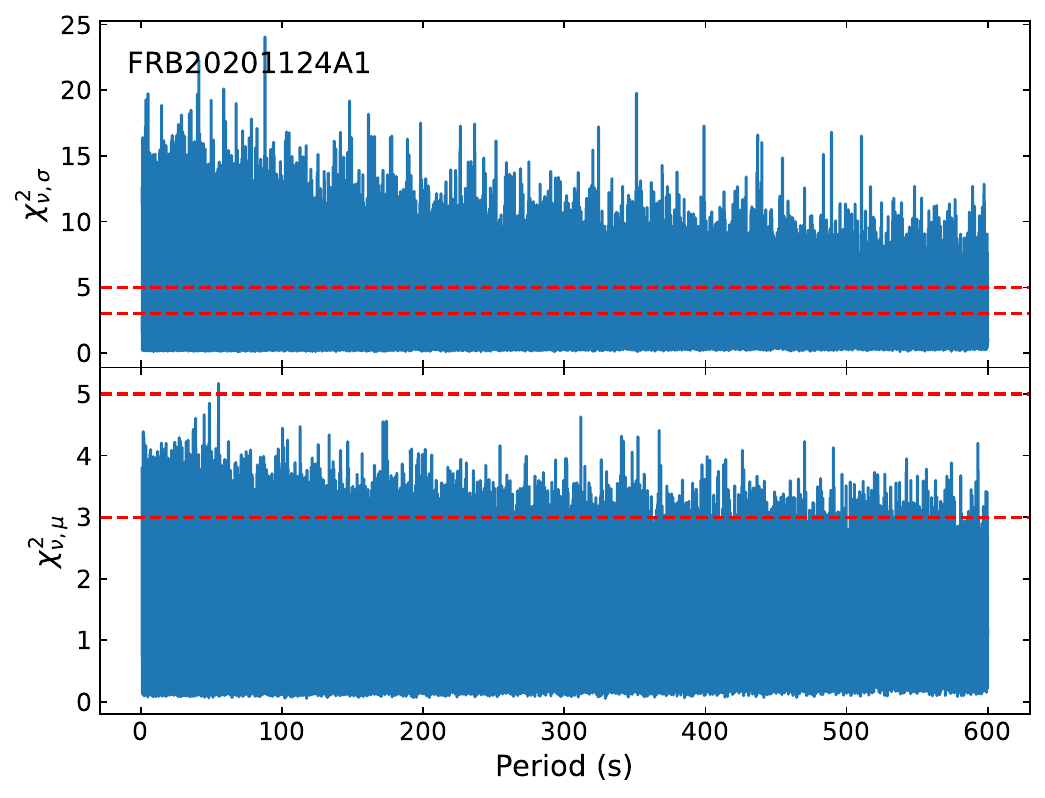}}
\subfigure{\includegraphics[width=3in]{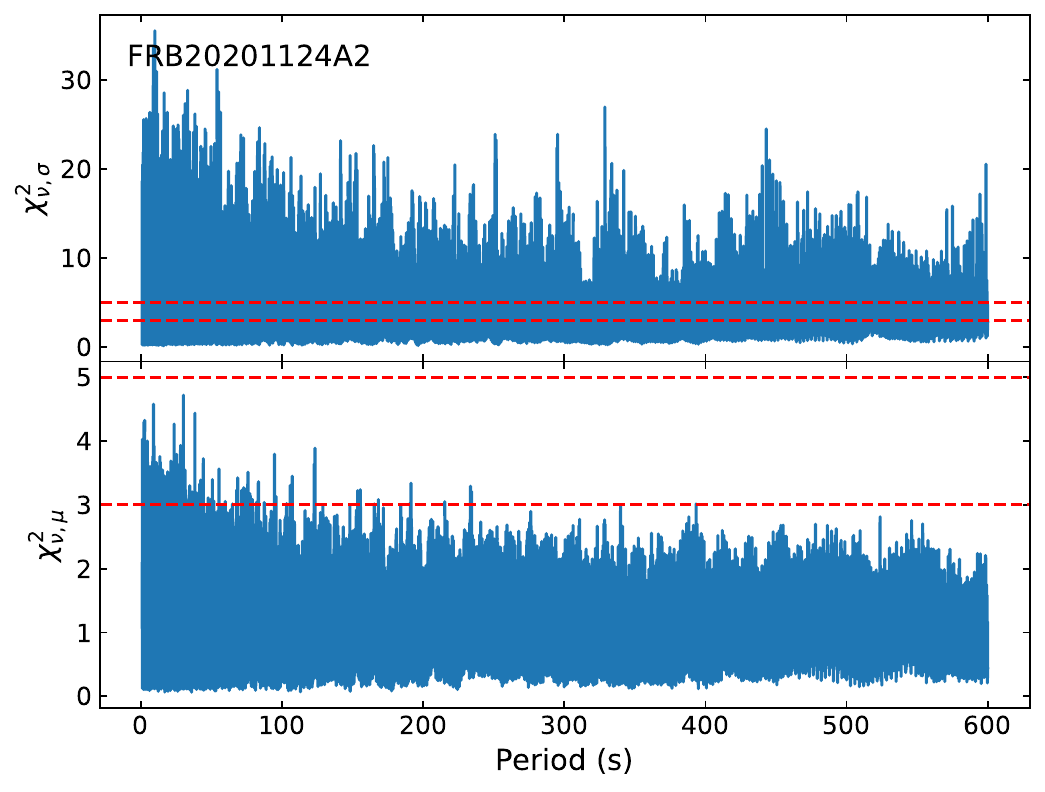}}
\subfigure{\includegraphics[width=3in]{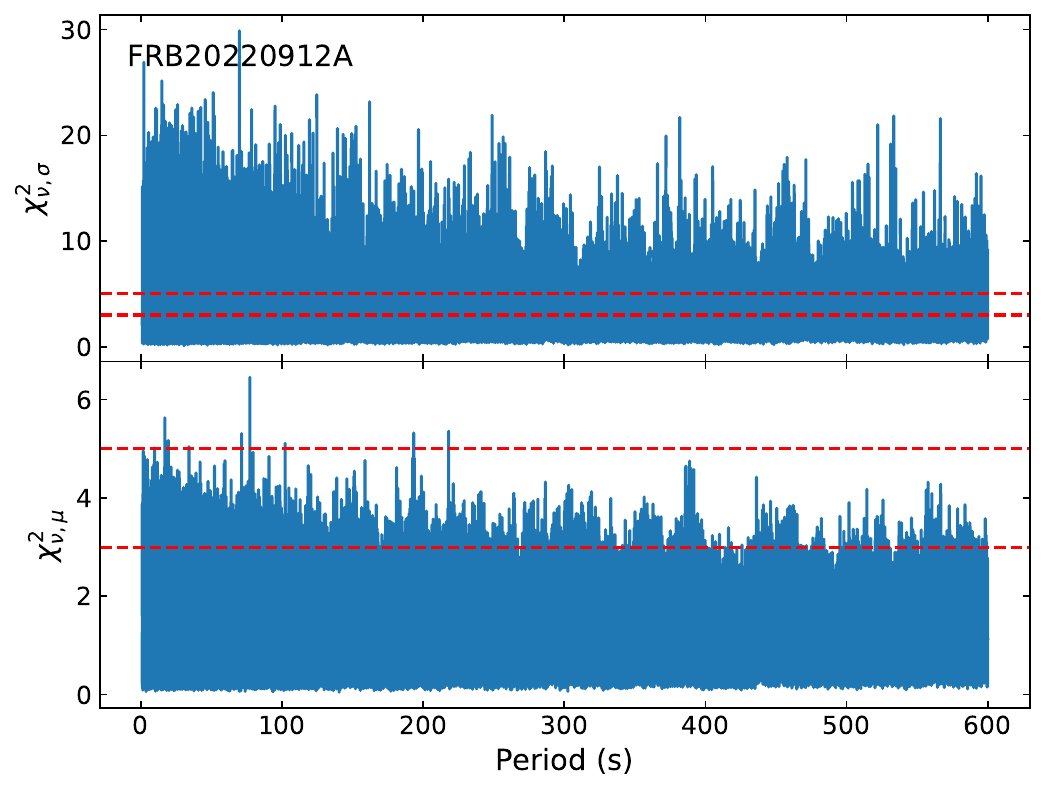}}
\subfigure{\includegraphics[width=3in]{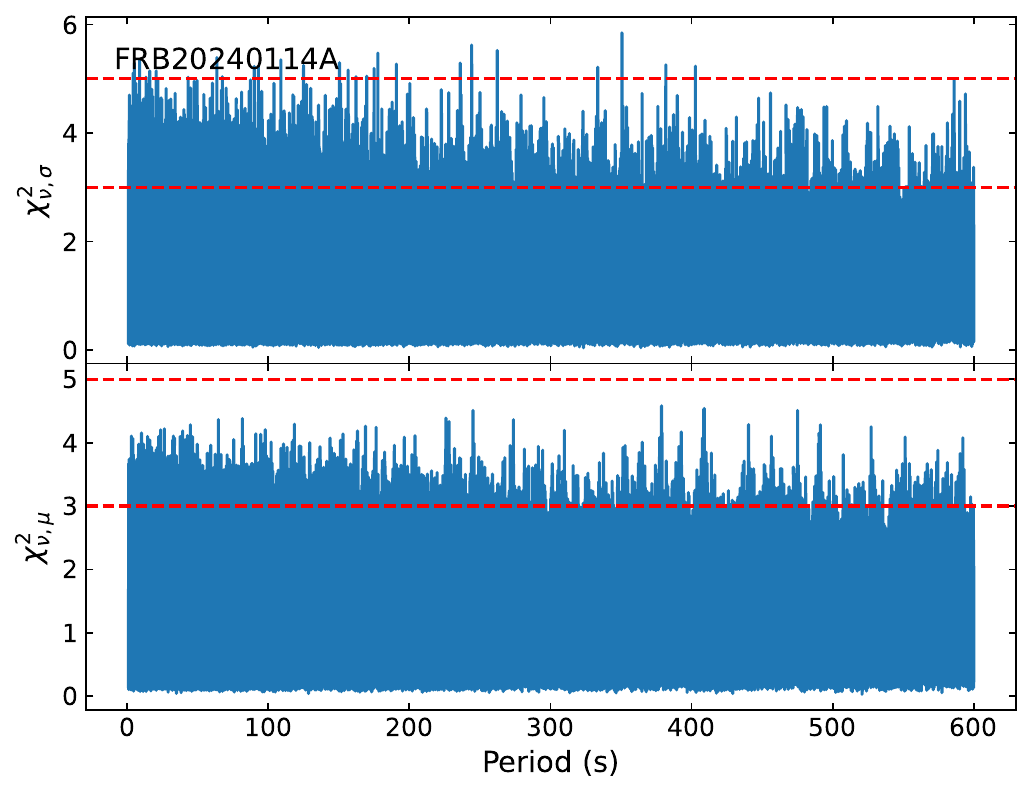}}
\caption{$\chi^2_{\nu,\mu}$ and $\chi^2_{\nu,\sigma}$ of all trial periods in the range $[1, 600]$ s for FRB 20201124A1, FRB 20201124A2, FRB 20220912A, and FRB 20240114A, respectively. The red dotted lines correspond to $\chi^2_{\nu}=3$ and $\chi^2_{\nu}=5$ following the criteria adopted earlier in the paper to characterize the strength of PA variation. These values do not correspond to formal statistical significance thresholds and are shown only as visual reference levels.}
\label{fig: realdata}
\end{figure*}

In Figure \ref{fig: realdata}, we show $\chi^2_{\nu,\mu}$ and $\chi^2_{\nu,\sigma}$ of periods. In all the panels, we do not observe a significant simultaneous rise in $\chi^2_{\nu,\mu}$ and $\chi^2_{\nu,\sigma}$, and the maximums of $\chi^2_{\nu,\mu}$ are usually comparable to the noise region, which suggests that no significant periodicity is detected. Though in the case of FRB 20220912A, the maximum of $\chi^2_{\nu,\mu}$ is larger than 5, but the corresponding $\chi^2_{\nu,\sigma}$ is small. To verify the authenticity of these candidates, we also check the burst number, $\chi^2_{\nu,\mu}$ and $\chi^2_{\nu,\sigma}$ of the binned phase, but no simultaneous trend is identified. These results suggest that the direction of the mean magnetic axis remains consistent at all times. This may be attributed to a stable geometric configuration or a highly aligned rotation and magnetic axes.

Since we find no evidence for a rotational scenario in the data, we attempt to fit the observed PA distribution using Model A. However, due to the strong degeneracy between the impact angle $\beta$ and the concentration parameter $\kappa$, fitting the observed PA distribution directly with Model A does not yield any meaningful constraints.

\section{Discussion} \label{sec: imp}
\subsection{Emission Mechanism}
PA provides a powerful diagnostic of the radiation mechanism of FRBs. In shock-related models, the highly ordered magnetic field can not easily account for either the rapid PA swings within bursts or PA variations from burst to burst \citep{2019MNRAS.485.4091M, 2020ApJ...896..142B}. On the other hand, the magnetospheric model can attribute PA variations to the geometric effect or magnetospheric configuration reshaping \citep{2017MNRAS.468.2726K, 2017ApJ...836L..32Z, 2018ApJ...868...31Y, 2022ApJ...927..105W, 2022ApJ...925...53Z, 2023MNRAS.522.2448Q, 2024ApJ...972..124Q}.
Given the upper limit of the randomized time scale inferred from SF method is shorter than 10 ms, the causality implies that the coherence scale should satisfy $L_{\mathrm{coh}} \lesssim c \tau \lesssim 3 \times 10^3 ~\mathrm{km}$ ($\lesssim$ 300 $R_{\mathrm{NS}}$), where $c$ is the speed of light, $\tau$ is the randomized time scale, and $R_{\mathrm{NS}} = 10 ~\mathrm{km}$ is the typical radius of the neutron star. If the observed PA is determined primarily by the local magnetic-field orientation near the neutron star surface, this scale may also provide an upper limit on the emission altitude.

The geometric effect further predicts that the PA behavior should exhibit periodic modulation tied to the spin of the source, even in the scenario of the near-aligned magnetic and rotational axes. The stable statistical properties of PA can also be explained by the near-aligned magnetic and rotational axes, which maintain a stable geometric orientation with respect to the observer.
In contrast, PA variations driven by the dynamically evolving magnetic configuration are intrinsically stochastic, as the underlying perturbations are not controllable. Such perturbations may originate from the distortion of the FRB emitter itself \citep{2021NatAs...5..378L, 2020ApJ...898L..29M}, and lead to transient offsets of the effective magnetic axis away from the center of the emission region for each burst, while the PA distribution remains statistically clustered around the emission-region center. This additional layer of dynamical evolution offers a natural explanation for why no clear periodicity is observed in the PA-based period search.

Moreover, the observed burst-to-burst PA variations closely resemble those seen in SGR J1935+2154 \citep{2023SciA....9F6198Z} and 1E 1547.0-5408 \citep{Lower:2026bch}, further supporting the picture that the magnetic axis does not maintain a stable orientation.
Such erratic polarization variations imply that the emission geometry is frequently restructured by intense magnetic activity, rather than being confined to a static lighthouse-like beam.
Together, these behaviors strengthen the phenomenological similarity between FRBs and magnetars and point to an intrinsically dynamical evolving magnetosphere.

\subsection{FRB Population}
Our model also provides a natural interpretation for the population difference between repeating and apparently non-repeating FRBs. In this picture, the emission is associated with a magnetic axis that undergoes burst-to-burst perturbations, forming a dynamically evolving magnetosphere. For most viewing geometries, such random perturbations only rarely bring the instantaneous emission direction into the LOS, so bursts may be detected only once or extremely infrequently. In this sense, apparently non-repeating FRBs can arise naturally from the dynamical RVM without requiring a special geometrical configuration. On the other hand, favorable geometries can still account for highly active repeaters \citep{2025ApJ...982...45B}. In addition, the dynamical evolution of the magnetic axis can smear out the strict phase-dependent PA behavior predicted by a standard rotating vector model, thereby reconciling geometric models with the widely scattered PA values observed in many repeating sources. Such a framework is also broadly compatible with rotation-modulated magnetar burst scenarios \citep{2025ApJ...988...62L}, in which burst activity is linked to rotation modulation but can appear observationally stochastic.

\section{Conclusion} \label{sec: con}
In this work, we conducted a systematic search for periodic behavior in the PA of repeating FRBs using LSP analyses of four large PA datasets observed by FAST from three active sources: FRB 20201124A, FRB 20220912A, and FRB 20240114A. These datasets span long time baselines and contain dense PA measurements, providing a unique opportunity to probe potential periodic signatures beyond individual burst durations. Although the PA exhibits significant burst-to-burst variations, no statistically significant periodicity is detected across the explored timescale ranges. Monte Carlo simulations demonstrate that the strongest peaks in the LSPs can be reproduced by the intrinsic sampling window of the burst sequence combined with internal burst structures, indicating that these features are primarily caused by aliasing rather than by the spin or orbital motion of the central engine.

To interpret these results, we extend the traditional RVM to a dynamically evolving version, in which stochastic perturbations allow the effective magnetic axis to vary from burst to burst. In this model, the emission geometry evolves around a mean magnetic configuration, producing the burst-to-burst PA variations observed in the data. Once such stochastic perturbations dominate the whole emission processes, periodic rotation is unlikely to appear directly in the PA time series. Instead, any underlying periodicity would be expected to manifest as phase-dependent modulations in the statistical properties of the PA population or in the burst occurrence rate. Motivated by these predictions, we searched for such signatures in both the PA population and the burst rate, but no statistically significant periodic modulation was detected, which suggests a stable geometry or a highly aligned rotation and magnetic axes.
The diversity of PA behaviors observed among repeating FRBs may therefore reflect different magnetospheric states or evolutionary stages. These results highlight the important role of magnetospheric dynamics in shaping the polarization properties of FRB emission and provide further evidence supporting magnetar-like central engines for repeating FRBs.

\begin{acknowledgments}
This work made use of data from the FAST FRB Key Science Project.
X.-H. L. acknowledges support from the National SKA Program of China （Nos. 2022SKA0110100 and 2022SKA0110101) and NSFC (grant No. 12361141814).
J.-R. N. is supported by the National Natural Science Foundation of China (NSFC, No. 12503055) and the Postdoctoral Fellowship Program of CPSF under Grant Number GZB20250737.
W.-Y. W. acknowledges support from the NSFC (No.12261141690 and No.12403058), the National SKA Program of China (No. 2020SKA0120100), and the Strategic Priority Research Program of the CAS (No. XDB0550300).
X.-L. C. acknowledges support from the NSFC (grant No. 12361141814, 12421003), and by the Specialized Research Fund for State Key Laboratory of Radio Astronomy and Technology.
D. L. acknowledges support from the New Cornerstone Foundation.
\end{acknowledgments}

%

\vspace{5mm}






\bibliography{sample701}{}
\bibliographystyle{aasjournal}


\end{CJK*}
\end{document}